\documentclass[journal]{IEEEtran}
\usepackage{amsmath,graphicx}
\usepackage{amsfonts}
\usepackage{amssymb}
\usepackage{xspace}
\usepackage{bm}
\usepackage{mathtools} 
\usepackage{hyperref}
\usepackage{soul}

\long\def\comment#1{}
 
\usepackage{xcolor}
\usepackage{xspace}
\newcommand{\xmath}[1] {\ensuremath{#1}\xspace}
\long\def\red#1{\bgroup\color{red}#1\xspace\egroup}
\long\def\blue#1{\bgroup\color{black}\ignorespaces#1\xspace\egroup}

\long\def\purple#1{\bgroup\color{purple}#1\xspace\egroup}

\newcommand{\fref}[1] {Fig.~\ref{#1}\xspace} 
\newcommand{\tref}[1] {Table~\ref{#1}\xspace}
\newcommand{\sref}[1] {\S\ref{#1}\xspace}
\newcommand{\resp}[1]{}

\newcommand{\blmath}[1] {\xmath{\bm{#1}}}
\newcommand{\Expni}[1] {\xmath{\mathrm{e}^{-\imath #1}}}
\newcommand{\Expri}[1] {\xmath{\mathrm{e}^{\imath #1}}}
\newcommand{\normii}[1] {\xmath{\left\| #1 \right\|_2}}
\newcommand{\normiir}[1] {\xmath{\| #1 \|_2}}
\newcommand{\norminf}[1] {\xmath{\left\| #1 \right\|_{\infty}}}
\newcommand{\normfro}[1] {\xmath{\left\| #1 \right\|_{\mathrm{F}}}}
\newcommand{\normfror}[1] {\xmath{\| #1 \|_{\mathrm{F}}}}
\newcommand{\norminfr}[1] {\xmath{\| #1 \|_{\infty}}}
\newcommand{\brak}[1] {\xmath{\left[ #1 \right]}}
\newcommand{\paren}[1] {\xmath{\left( #1 \right)}}
\newcommand{\tagl}[1]{\tag{#1}\label{#1}}
\newcommand{\bset}[1]{\xmath{\left\{#1\right\}}}

\newcommand{\omdd}[1] {\xmath{\bm{\omega}^{[#1]}}}
\newcommand{\omd} {\omdd{d}}
\newcommand{\omde} {\omega^{[d]}} 
\newcommand{\omdl} {\xmath{\omde_l}}
\newcommand{\rd} {\xmath{\bm{r}^{[d]}}}
\newcommand{\vect} {\operatorname{vec}}
\newcommand{\bmat}[1] {\xmath{\begin{bmatrix} #1 \end{bmatrix}}}

\newcommand{\gmax} {\xmath{g_{\max}}}
\newcommand{\smax} {\xmath{s_{\max}}}

\newcommand{\omi} {\xmath{\vec{\omega}_i}}

\newcommand{\omgm} {\xmath{\omega}_m}
\newcommand{\oml} {\xmath{\vec{\omega}_l}}
\newcommand{\xj} {\xmath{x_j}}
\newcommand{\yi} {\xmath{y_i}}
\newcommand{\rr} {\xmath{\blmath{r}}}
\newcommand{\rj} {\xmath{\vec{r}_j}}
\newcommand{\rdj} {\xmath{r^{[d]}_j}}
\newcommand{\rdk} {\xmath{r^{[d]}_k}}
\newcommand{\rk} {\xmath{\vec{r}_k}}
\newcommand{\aij} {\xmath{a_{ij}}}
\newcommand{\amn} {\xmath{a_{mn}}}
\newcommand{\emn} {\xmath{e_{mn}}}
\newcommand{\amnt} {\xmath{\tilde{a}_{mn}}}
\newcommand{\wj} {\xmath{w_j}}
\newcommand{\ti} {\xmath{t_i}}
\newcommand{\Nc} {\xmath{N_{\mathrm{c}}}}
\newcommand{\veps}{\xmath{\varepsilon}}

\newcommand{\defequ} {\triangleq}

\newcommand{\diag}[1] {\mathrm{diag}\!\left\{#1\right\}}
\newcommand{\real}[1] {\mathrm{real}\!\left\{#1\right\}}

\newcommand{\abs}[1] {\left| #1 \right|}

\newcommand{\JM}[2] {\xmath{\mathcal{D}_{#2} \, #1}}
\newcommand{\trace}[1] {\mathrm{Tr}\!\left\{#1\right\}}
\newcommand{\Epx}[1] {\mathsf{E}_{p(\x)}\![ #1 ]}

\newcommand{\reals} {\xmath{\mathbb{R}}}
\newcommand{\complex} {\xmath{\mathbb{C}}}

\newcommand{\RNN} {\xmath{\reals^{N \times N}}}
\newcommand{\CN} {\xmath{\complex^N}}
\newcommand{\CM} {\xmath{\complex^M}}
\newcommand{\CMN} {\xmath{\complex^{M \times N}}}
\newcommand{\CNN} {\xmath{\complex^{N \times N}}}

\renewcommand{\a} {\blmath{a}}

\newcommand{\e} {\blmath{e}}
\newcommand{\A} {\blmath{A}}
\newcommand{\Anu} {\blmath{\tilde{A}}}
\newcommand{\B} {\blmath{B}}
\newcommand{\C} {\blmath{C}}
\newcommand{\D} {\blmath{D}}
\newcommand{\E} {\blmath{E}}
\newcommand{\F} {\blmath{F}}
\newcommand{\Ft} {(\blmath{F}^T)}
\newcommand{\G} {\blmath{G}}
\newcommand{\Gt} {(\blmath{G}^T)}

\newcommand{\I} {\blmath{I}}
\newcommand{\J} {\blmath{J}}

\newcommand{\Jnu} {\blmath{\tilde{J}}}
\newcommand{\T} {\blmath{T}}
\newcommand{\R} {\mathsf{R}} 
\newcommand{\Sm}{\blmath{S}}
\newcommand{\Em}{\blmath{E}}
\newcommand{\Ef}{\blmath{E}_{\mathrm{f}}}
\newcommand{\W} {\blmath{W}}

\newcommand{\Xm} {\blmath{X}}
\newcommand{\x} {\blmath{x}}
\newcommand{\z} {\blmath{z}}
\newcommand{\Y} {\blmath{Y}}

\newcommand{\xk} {\xmath{\blmath{x}_k}}
\newcommand{\xtrue} {\xmath{\blmath{x}^{\mathrm{true}}}}
\newcommand{\xkk} {\xmath{\blmath{x}_{k+1}}}
\newcommand{\y} {\blmath{y}}

\renewcommand{\v} {\blmath{v}}
\newcommand{\vveps} {\blmath{\varepsilon}}
\newcommand{\om} {\blmath{\omega}}
\newcommand{\s} {\blmath{s}}

\newcommand{\dL} {\partial L}
\newcommand{\dz} {\partial \z}
\newcommand{\ds} {\partial \s}

\usepackage{slashbox}

\begin{document}

\title{Efficient Approximation of Jacobian Matrices
Involving a Non-Uniform Fast Fourier Transform (NUFFT)}
%
\author{Guanhua Wang and Jeffrey A. Fessler%
\thanks{This work supported in part by
NIH Grants R01 EB023618 and U01 EB026977
and NSF Grant IIS 1838179.}
\thanks{G. Wang is with the
Department of Biomedical Engineering, University of Michigan, Ann Arbor, MI 48109 USA (e-mail:guanhuaw@umich.edu).}
\thanks{J. A. Fessler is with the Department of EECS, University of Michigan, Ann Arbor, MI 48109 USA (e-mail:fessler@umich.edu).}
}

\maketitle
\begin{abstract}
%
There is growing interest in learning Fourier domain sampling strategies (particularly for magnetic resonance imaging, MRI)
using optimization approaches.
For non-Cartesian sampling,
the system models typically involve
non-uniform fast Fourier transform (NUFFT) operations.
Commonly used NUFFT algorithms contain frequency domain interpolation,
which is not differentiable
with respect to the sampling pattern,
complicating the use of gradient methods.
This paper describes an efficient and accurate approach
for computing approximate gradients
involving NUFFTs.
Multiple numerical experiments validate
the improved accuracy and 
efficiency of the proposed approximation.
As an application to computational imaging, 
the NUFFT Jacobians were used to
optimize non-Cartesian MRI sampling trajectories
via data-driven stochastic optimization.
Specifically, the sampling patterns were learned 
with respect to
various model-based image reconstruction (MBIR) algorithms.
The proposed approach enables
sampling optimization for
image sizes
that are infeasible
with standard auto-differentiation methods
due to memory limits.
The synergistic acquisition and reconstruction design
leads to remarkably improved image quality.
In fact,
we show that
model-based image reconstruction methods
with suitably optimized imaging parameters
can perform nearly as well as CNN-based methods.

\end{abstract}
\section{Introduction}
\label{sec:intro}

There are several computational imaging modalities
where the raw measurements
can be modeled as
samples
of the imaged object's spectrum,
where those samples need not
lie on the Cartesian grid,
including
radar
\cite{munson:84:irf},
diffraction ultrasound tomography
\cite{bronstein:02:rid},
parallel-beam tomography
\cite{matej:04:iti},
and MRI
\cite{wright:14:ncr,fessler:10:mbi}.
Image reconstruction methods for such modalities
may use
non-uniform fast Fourier transform (NUFFT) operations
to accelerate computation
\cite{fessler:03:nff,yang:14:mso}.
The quality of the reconstructed image
depends both on the image reconstruction method
and on the characteristics
of the frequency domain sampling pattern.

\resp{R1.1}
MRI has particular flexibility in designing frequency domain sampling patterns.
\blue{
Many MR sampling patterns are discrete subsets of the Cartesian grid,
and the corresponding optimization/learning strategies
include
greedy algorithms
\cite{zibetti2020fast,sanchez:2020:ScalableLearningBasedSampling,gozcu:2019:RSP},
reparameterization
\cite{cao:1993:Feature,knoll:11:ars,bahadir:2020:LOUPE,sherry:20:lts, huijben:2020:LearningSamplingModelBased},
Bayesian optimization
\cite{seeger:2010:OptimizationKspaceTrajectories,haldar:2019:OEDIPUS},
and system matrix analysis
\cite{vonkienlin:1991:Spectral,gao:2000:OptimalKspaceSampling,xu2005optimal, levine2017fly}.
The other type is non-Cartesian sampling,
which uses
a collection of continuous functions in k-space.
Several studies applied gradient
methods to optimize non-Cartesian sampling trajectories
\cite{pilot,aggarwal:20:jmj,wang:22:bjork-tmi,scope2022efficient},
and it is also possible to use
derivative-free optimization algorithms
in certain applications \cite{jordan2021automated}.}
This paper develops efficient tools
for applying gradient methods
to non-Cartesian sampling pattern optimization.

Some data-driven optimization methods for non-Cartesian sampling 
solve an optimization problem
involving both forward system models
and image reconstruction methods \cite{pilot,aggarwal:20:jmj,wang:22:bjork-tmi}.
The forward models and reconstruction methods both depend on NUFFT operations.
In principle, the Fourier transform operation 
is a continuous function
of the k-space sample locations
and thus should be applicable
to gradient-based optimization methods.
In practice,
the NUFFT
($\mathcal{O}(N \log N)$ operations)
is an approximation
to the non-uniform discrete Fourier transform
(NUDFT, $\mathcal{O}(N^2)$ operations)
and that approximation
often is implemented
using non-differentiable lookup table operations or other interpolation techniques
\cite{dale:01:arl,beatty:05:rgr}.
Such approximations are sufficient for image reconstruction (forward mode),
but have problematic efficiency and accuracy
if one attempts to use standard auto-differentiation tools
for gradient-based optimization.
Standard auto-differentiation methods
using subgradients
can lead to incorrect NUFFT Jacobians.
They also require prohibitively large amounts of memory
for back-propagation through certain algorithm stages
such as 
conjugate gradient (CG) steps
that involve NUFFT operations.

This paper proposes an efficient approach
that replaces
memory-intensive and inaccurate
auto-differentiation steps
with fast Jacobian approximations
that are themselves based on NUFFT operations.
The proposed approach
requires substantially less memory for iterative updates
like CG steps.

As a direct application,
we used the proposed Jacobian to learn
MRI sampling trajectories via 
stochastic optimization.
By applying the forward system model
and subsequent reconstruction,
reconstructed images 
were simulated from reference images
in the training set.
The similarity  
between simulated and reference images
was the metric for updating the sampling trajectory.
We used model-based reconstruction methods,
such as regularized least-squares and compressed sensing.
In comparison with
previous works using reconstruction neural networks (NN) \cite{pilot,wang:22:bjork-tmi},
such model-based reconstruction methods
can be more robust
and require less training data.

In addition to simple NUFFT-based sensing matrices, 
we also considered several scenarios in MR sampling
and reconstruction,
including the multi-coil (sensitivity-encoded) imaging
\cite{pruessmann_sense_1999}
system models that account for 
$B_0$ field inhomogeneity
\cite{sutton:2003:FastIterativeImagea}.
The derivation also includes fast Jacobian approximations
for Gram and ``data consistency" operations
commonly used in iterative reconstruction methods.

Jacobians with respect to the non-Cartesian sampling pattern
are also relevant
for tomographic image reconstruction problems
with unknown view angles
(like cryo-EM)
where the view angles must be estimated
\cite{zehni:20:jar}.

The remainder of this paper is organized as follows.
Section~\ref{s,jacob}
derives the efficient Jacobian approximations.
Section~\ref{s,opt}
details how to optimize MRI sampling patterns
using learning-based methods.
Section~\ref{sec:vali}
provides empirical validation of the approach,
showing the efficacy and accuracy of
the proposed approach.
The appendix includes an error analysis
of the proposed method.

The methods in this paper were used
to assist the design of k-space sampling
for a CNN-based reconstruction approach
in our previous work
\cite{wang:22:bjork-tmi}.
This paper derives the theory
in detail
and considers k-space sampling optimization
for general model-based reconstruction methods.
Preliminary results were shown in an earlier short conference  abstract
\cite{wang:22:rmb}.

\section{Jacobian Expressions}
\label{s,jacob}

This section derives the key Jacobian expressions
and their efficient approximations
based on NUFFT operations.
These approximations
enable the applications that follow.

\subsection{Lemmas}
We denote matrices, vectors and scalars by \A, \a and $a$, respectively.
$\A'$, $\A^{T}$ and $\A^*$
denote the Hermitian transpose,
the transpose and the complex conjugate of \A, respectively.

Consider a scalar function
$f(z),\ z = x + y\imath \in \complex,\ x,y \in \reals$.
Following the conventions in Wirtinger calculus
\cite[p.~67]{remmert1991theory},
the differential operators are defined as
\[
\frac{\partial}{\partial z} =
\frac{1}{2} \frac{\partial}{\partial x} - \frac{\imath}{2}  \frac{\partial}{\partial y},\
\frac{\partial}{\partial z^*} =
\frac{1}{2} \frac{\partial}{\partial x} + \frac{\imath}{2}  \frac{\partial}{\partial y}.
\]
A function
$f$ is \textit{complex differentiable} or \textit{holomorphic} iff
$\frac{\partial f}{\partial z^*} = 0$
(Cauchy–Riemann equation)
\cite[p.~66]{remmert1991theory}.
In the context of optimization,
a cost function
$L$
(usually a real scalar)
is not holomorphic w.r.t. complex variables.
A common approach (as adopted by PyTorch and TensorFlow)
regards the real and imaginary components
of a complex variable as two real-valued variables, 
and updates them separately,
similar to the real-valued calculus
\cite{kreutz2009complex}.
For example,
the $n$th gradient descent step
uses the update
\begin{equation*}
\z_{n+1} = \z_{n} - \alpha \paren{\frac{\dL}{\partial \x}
+ \imath \frac{\dL}{\partial \y} }
= \z_n - 2 \alpha \frac{\dL}{\partial \z^*}
,
\end{equation*}
where $\alpha \in \reals^+$
denotes the step size.
The chain rule still applies to calculating
$\frac{\dL}{\partial \z^*}$
\cite{hjorungnes2007complex}
\cite[p.~68]{remmert1991theory};
for $s = f(z)$:
\begin{equation}
\frac{\dL}{\dz^*}
= \paren{\frac{\dL}{\ds^*}}^* \frac{\ds}{\dz^*}
+ \frac{\dL}{\ds^*} \paren{\frac{\ds}{\dz}}^*.
\label{eqn:wir_chain}
\end{equation}

For Jacobian matrices,
we follow the
\href
{https://en.wikipedia.org/wiki/Matrix_calculus#Numerator-layout_notation}
{``numerator-layout''}
notation
\cite{wiki:numerator}.
For example,
the derivative of an $m$-element column vector \y
w.r.t. an $n$-element vector \x is
an $m \times n$ matrix:
\begin{equation}
\frac{\partial \y}{\partial \x}
\defequ
\left[\begin{array}{cccc}
\frac{\partial y_{1}}{\partial x_{1}} & \frac{\partial y_{1}}{\partial x_{2}} & \cdots & \frac{\partial y_{1}}{\partial x_{n}} \\
\frac{\partial y_{2}}{\partial x_{1}} & \frac{\partial y_{2}}{\partial x_{2}} & \cdots & \frac{\partial y_{2}}{\partial x_{n}} \\
\vdots & \vdots & \ddots & \vdots \\
\frac{\partial y_{m}}{\partial x_{1}} & \frac{\partial y_{m}}{\partial x_{2}} & \cdots & \frac{\partial y_{m}}{\partial x_{n}}
\end{array}\right].
\label{eq:dydx}
\end{equation}
However,
this convention does not handle scenarios
such as the derivatives of the elements of one matrix
w.r.t. the elements of another matrix.
Thus, we adopt a natural extension
by using the $\vect$ (vectorization) operation.
Specifically,
for a $M\times N$ matrix \A
that is a function
of a $P \times Q$ matrix \B,
we write the derivative as a $MN \times PQ$ matrix
by applying
\eqref{eq:dydx}
to the $\vect$ of each matrix:
\begin{equation}
\JM{\A}{\B}
= \JM{\A(\B)}{\B}
\defequ
\frac{\partial \vect(\A)}{\partial \vect(\B)}.
\label{e,D}
\end{equation}

The following equalities are useful in our derivations.
(Equalities involving products
all assume the sizes are compatible.)
\comment{
\begin{align*}
   \text{(P1)} \qquad \vect (\A \B \C)&=(\I_{n} \otimes \A \B) \vect(\C) \\
   &=(\C^T\B^T \otimes \I_{k}) \vect(\A) 
\end{align*}
\[
\text{(P2)} \qquad
(\A \otimes \B) (\C \otimes \D) = (\A \C) \otimes(\B \D)
\]
\[
\text{(P3)} \quad 
\A \otimes \B=(\I \otimes \B)(\A \otimes \I)=(\A \otimes \I)(\I \otimes \B)
\]
P1-P3 are common matrix vectorization conclusions.
\[
\text{(P4)} \qquad
\JM{\A\x}{\A} = \x^{T}\otimes\I
\]
\[
\text{(P5)} \qquad
\A \in \RNN, \, 
\JM{\A^{-1}}{A} = -(\A^T)^{-1} \otimes \A^{-1}.
\]
}
For
\(
\A \in \complex^{K \times L}
,\
\B \in \complex^{L \times M}
,\
\C \in \complex^{M \times N}
\):
\begin{align}
\vect (\A \B \C) &= (\I_{N} \otimes \A \B) \vect(\C)
   \nonumber
   \\&=
   (\C^T\B^T \otimes \I_{K}) \vect(\A).
\tagl{P1}
\end{align}
In general:
\begin{equation}
    (\A \otimes \B)(\C \otimes \D)=(\A \C) \otimes(\B \D).
\tagl{P2}
\end{equation}
For
\(
\A \in \complex^{K \times L}
,\
\B \in \complex^{M \times N}
\):
\begin{equation}
    \A \otimes \B=(\I_{K} \otimes \B)(\A \otimes \I_{N})=(\A \otimes \I_{M})(\I_{L} \otimes \B).
\tagl{P3}
\end{equation}
For
\(
\A \in \complex^{M \times N}
,\
\x \in \complex^{N}
\):
\begin{equation}
    \JM{ (\A\x)}{\A} = \x^T \otimes \I_M,~
    \JM{ (\A\x)}{\A^*} = \blmath{0}
\tagl{P4}
.\end{equation}
For an invertible matrix \A:
\begin{align}
\A \in \CNN \implies &
\JM{\A^{-1}}{\A} = -(\A^T)^{-1} \otimes \A^{-1},
\nonumber \\
& \JM{\A^{-1}}{\A^*} = \blmath{0}
.
\tagl{P5}
\end{align}
The chain rule still holds
for the extended Jacobian formulation.
Suppose
$F : \, \complex^{K \times L} \to \complex^{M \times N}$
and 
$G : \, \complex^{M \times N} \to \complex^{P \times Q}$ are both holomorphic.
For $\Xm \in \complex^{K \times L}$,
the Jacobian of the composite function is:
\begin{align}
& \underbrace{
    \JM{G(F(\Xm))}{\Xm}
}_{PQ \times KL}    
=
\underbrace{
    \left. \JM{G(\Y)}{\Y} \right|_{\Y = F(\Xm)}
}_{PQ \times MN}
    \,
\underbrace{
    \JM{F(\Xm)}{\Xm}
}_{MN \times KL} \ , \nonumber \\
&
\JM{G(F(\Xm))}{\Xm^*} = \blmath{0}.
\tagl{P6}
\end{align}
Equalities
\eqref{P1}-\eqref{P3}
are common matrix vectorization properties.
See \cite[Ch.~9]{magnus2019matrix}
for \eqref{P4}, 
\cite{hjorungnes2007complex} for \eqref{P5} and \eqref{P6}.

\subsection{System Model}
\label{sec,model}

Consider the (single-coil, initially) MRI measurement model
for non-Cartesian sampling
based on the
NUDFT
\cite{fessler:10:mbi}:
\[
\y = \A \x + \vveps
,\]
where
$\y \in \CM$
denotes the measured k-space data,
$\x \in \CN$
denotes the unknown image
to be reconstructed,
and
$\A \in \CMN$
denotes the system matrix
or encoding matrix,
where
$\A = \A(\om)$
has elements
\begin{equation}
    \aij = \Expni{\omi \cdot \rj}
,\quad
i=1,\ldots, M
,\quad
j=1,\ldots,N
\label{e,ndft}
\end{equation}
for
$\omi \in \reals^D$
and
$\rj \in \reals^D$
where
$D \in \{1,2,3\ldots\}$
denotes the image dimension,
and where
\[
\om = [\omdd{1} \ \omdd{2} \ldots \omdd{D}]
\]
is the $M \times d$ matrix
consisting of all the k-space sampling locations
and
$\omd \in \reals^M$
denotes its $d$th column.
(For simplicity here,
we ignore other physical effects
like
field inhomogeneity and relaxation
that are sometimes included
in the forward model in MRI
\cite{fessler:10:mbi}.)
The center locations of voxels
\bset{\rj}
usually lie on a Cartesian grid,
but the k-space sample locations \om
in principle can be arbitrary
subject to the Nyquist constraint.

Typically \A is approximated
by a NUFFT
\cite{fessler:03:nff}.
Usually, the NUFFT operator involves
frequency-domain interpolation operations
that are often non-differentiable.
One previous trajectory optimization approach
that used auto-differentiation \cite{pilot}
replaced the non-differentiable lookup table
with a bilinear interpolator.
Bilinear interpolation is differentiable
everywhere except at the sample locations.
Auto-differentiation of bilinear interpolation
involves differentiating
some \texttt{floor} and \texttt{ceiling} operations
and those derivatives are defined to be zero
in popular deep learning
frameworks such as PyTorch and TensorFlow,
leading to
suboptimal sub-gradient calculations.
Nearest-neighbor interpolation
has even worse properties
for auto-differentiation
because its derivative is zero almost everywhere,
leading to a completely vanishing gradient.

In the following derivations,
we investigate a different approach
where we analyze
the Jacobians
w.r.t. \om and \x
using the NUDFT expression \eqref{e,ndft}.
Then for efficient implementation,
we replace the NUDFT operations within the Jacobians with NUFFT approximations.
This approach enables faster computation and requires substantially less memory.

\subsection{Forward Operator}
\label{sec,for}

We first focus on the forward operation
$\A(\om) \, \x$
and determine Jacobian matrices
with respect to \x and \om.
The $M \times N$
Jacobian matrix of the forward linear operation
with respect to \x is
\[
\frac{\partial \A\x}{\partial \x} = \A,
\qquad
\frac{\partial \A\x}{\partial \x^*} = \blmath{0}
.\]
For the $d$th column
of the spectrum sampling pattern \om,
the Jacobian has elements
\begin{align*}
\brak{ \frac{\partial \A\x}{\partial \omd} }_{il} & = 
\frac{\partial [\A\x]_i}{\partial \omdl} 
= \frac{\partial}{\partial \omdl} 
\sum_{j=1}^N \Expni{\omi \cdot \rj} \xj \\
& = \begin{cases}
       -\imath \sum_{j=1}^N \Expni{\omi \cdot \rj} \xj \rdj, & i = l \\
      0, & \text{otherwise,}
    \end{cases}
\end{align*}
for $i,l = 1, \ldots, M$.
The above summation is the product
of the $i$th row of
$-\imath \A$
with $\x \odot \rd$.
Thus the $M \times M$ Jacobian matrix
for the partial derivatives of $\A\x$
w.r.t. \omd is:
\begin{equation}
\frac{\partial \A\x}{\partial \omd} = -\imath \, \diag{\A (\x\odot\rd)}
\label{e,Ax}
.\end{equation}
Consequently, the Jacobian calculation
should apply \A to vector $\x\odot\rd$ once.
In the above derivation,
\A is a NUDFT operator.
In the practical implementation,
we use a NUFFT
to approximate \A,
both for the forward model
and for the Jacobian calculation.

\subsection{Adjoint Operator}
\label{sec,adj}

Derivations of the Jacobians for the adjoint operation
\(
\A'(\om) \, \y
\)
follow a similar approach.
For \y:
\[
\frac{\partial \A'\y}{\partial \y} = \A',
\qquad
\frac{\partial \A'\y}{\partial \y^*} = \blmath{0}
.\]
For the $d$th column of \om,
the $N \times M$ Jacobian matrix has elements:
\begin{align*}
\brak{ \frac{\partial \A'\y}{\partial \omd} }_{jl}
& = 
\frac{\partial [\A'\y]_j}{\partial \omdl}
= \frac{\partial \sum_{i=1}^M \Expri{\omi \cdot \rj} \yi}{\partial \omdl}
\\&
= \imath \Expri{\omi \cdot \rj} \yi \rdj
.\end{align*}
Thus the Jacobian matrix is
\begin{equation}
\frac{\partial \A'\y}{\partial \omd} =
\imath \, \diag{\rd} \A' \diag{\y} 
\label{e,A'y}
.\end{equation}

\subsection{Gram Matrix}
\label{sec,gram}

The product $\A'(\om) \, \A(\om) \, \x$
of the Gram matrix
of the NUDFT
with a vector
also arises in optimization steps
and requires appropriate Jacobian matrices.
For \x:
\[
\frac{\partial \A'\A \x}{\partial \x} = \A'\A,
\qquad
\frac{\partial \A'\A \x}{\partial \x^*} = \blmath{0}.
\]
The $(k,j)$th
element of the $N \times N$ matrix
containing the partial derivatives
of the Gram matrix
w.r.t. \omdl
is
\begin{align}
\left[ \frac{\partial \A'\A}{\partial \omdl} \right]_{k,j}
&=
\frac{\partial}{\partial \omdl}
\sum_{i=1}^M \Expni{\omi \cdot (\rj -\rk)} \nonumber
\\&=
-\imath \, (\rdj - \rdk) \, \Expni{\oml \cdot (\rj -\rk)} \nonumber
\\&=
-\imath \, (\rdj - \rdk) \, a^*_{lk} a_{lj}
.\end{align}
In matrix form:
\begin{align}
 \frac{\partial \A'\A}{\partial \omdl} 
 & = 
 \imath \, \diag{\rd} \A' \e_l \e_l' \A
- \imath \A' \e_l \e_l' \A \diag{\rd} 
\label{e,A'A,omdl}
.\end{align}
When multiplying the Jacobian with a vector \x:
\begin{align}
\frac{\partial \A'\A}{\partial \omdl} \x
&=\imath \, \diag{\rd} \a_l (\a_l' \x)
- \imath \a_l \a_l' \diag{\rd} \x \nonumber
\\
& = \imath \, (\a_l'\x) (\rd \odot \a_l) - \imath ( \a_l' (\x \odot \rd) ) \a_l
\label{e,A'A,x,omdl}
,\end{align}
where $\a_l = \A' \e_l$
denotes the $l$th column of $\A'$.


Consider the extended Jacobian expression:
\[
\JM{\A'\A}{\omdl} = \vect\paren{ \frac{\partial \A'\A}{\partial \omdl} }
.\]
Multiplying by \x yields:
\begin{align*}
\frac{\partial \A'\A}{\partial \omdl} \x &= 
\vect \paren{ \frac{\partial \A'\A}{\partial \omdl} \x } \\
&= (\x^T \otimes \I_N) \vect \paren{ \frac{\partial \A'\A}{\partial \omdl} } && \text{(use P1)}\\
&= (\x^T \otimes \I_N) \, \paren{ \JM{\A'\A}{\omdl} } \\
&= \paren{ \JM{\A'\A\x}{\A'\A} } \paren{ \JM{\A'\A}{\omdl} } && \text{(use P4)} \\
&= \JM{\A'\A\x}{\omdl}. && \text{(use P6)}
\end{align*}
Concatenating \eqref{e,A'A,x,omdl} by columns
leads to
the matrix
\begin{align}
\MoveEqLeft{
\bmat{ \dfrac{\partial \A'\A}{\partial \omde_1}
& \ldots &
\dfrac{\partial \A'\A}{\partial \omde_M} } \x
}
=
-\imath \, \A' \diag{\A (\x \odot \rd)}
\nonumber \\&
\hspace*{7em}
+ \imath \, \diag{\rd} \A' \diag{\A\x}.
\label{e,A'A,x,omd}
\end{align}
Alternatively, 
we can express the extended Jacobian as
\begin{align}
\MoveEqLeft{
\bmat{ \dfrac{\partial \A'\A}{\partial \omde_1}
     & \ldots &
     \dfrac{\partial \A'\A}{\partial \omde_M} } \x
}
\nonumber \\
     &= (\x^T \otimes \I_n) \paren{\JM {\A'\A}{\omd}} \nonumber \\
     &= \paren{ \JM{\A'\A\x}{\A'\A} } \paren{ \JM{\A'\A}{\omd} } \nonumber \\
     &= \JM{\A'\A\x}{\omd}.
    \label{D,A'A,x,omd}
\end{align}
Again we use NUFFT operations
to efficiently approximate \eqref{D,A'A,x,omd}.

\subsection{Inverse of Positive Semidefinite (PSD) Matrix}
\label{sec,inv}

Image reconstruction methods
based on algorithms
like the augmented Lagrangian approach
\cite{hestenes:1969:MultiplierGradientMethodsa}
use
``data consistency'' steps
\cite{aggarwal:19:mmb,ramani:11:pmi,chan:17:pap}
that often involve least-squares problems
with solutions in the following form:
\[
(\A'\A + \lambda \I)^{-1} \x,
\]
for some vector
$\x \in \C^N$,
or
\begin{equation}
(\A'\A + \lambda \T'\T)^{-1} \x,
\label{e,aatt,inv}
\end{equation}
where \T denotes a linear regularization operator
that is independent of \om.
In both cases,
$\lambda>0$
and the null spaces of \T and \A are disjoint,
so the Hessian matrix is invertible.
A few iterations of a CG method
usually suffices
to efficiently
compute the approximate product of such a matrix inverse
with a vector.
The direct inverse is impractical for large-scale problems
like MRI.
Following
\cite{aggarwal:19:mmb},
we treat CG as solving the above equations accurately,
so that we can derive efficient approximations as follows.
Otherwise, 
attempting to auto-differentiate
through a finite number of CG iterations
would require large amounts of memory.
Here we derive the corresponding Jacobian matrices
for the exact inverse
to \eqref{e,aatt,inv}
and then apply fast approximations.
For \x,
the $N \times N$ Jacobian is
\begin{align*}
& \frac{\partial(\A'\A + \lambda \T'\T)^{-1}\x}{\partial \x} = (\A'\A + \lambda \T'\T)^{-1}, \\
& \frac{\partial(\A'\A + \lambda \T'\T)^{-1}\x}{\partial \x^*} = 0.
\end{align*}

We can still use CG (with NUFFT)
to efficiently multiply this Jacobian
by a vector,
albeit approximately.

\newcommand{\eqntrick}{ \makebox[0pt][l]{ $-\imath \, \A' \diag{\A (\z \odot \rd)} $ }} 

To consider the Jacobian w.r.t.
the sampling pattern \omd,
define
$\z = (\A'\A + \lambda \T'\T)^{-1}\x$
and
$\F = \A'\A + \lambda \T'\T$.
We assume that \A and \T have disjoint null spaces,
so that \F is positive definite
and hence invertible.
Applying equalities derived above
leads to the following
expression
for the $M \times N$
Jacobian:
\begin{align}
\MoveEqLeft{
\JM {\F^{-1}\x}{\omd}
}
\nonumber \\
& = \paren{ \JM{\F^{-1}\x}{\F} } \paren{ \JM{\F}{\omd} }  \nonumber && \text{use P6} \\
& = -(\x^T \otimes \I) (\Ft^{-1} \otimes \F^{-1}) \paren{ \JM{\F}{\omd} }
&& \text{use P5}
\nonumber\\
& = - \paren{ (\x^T\Ft^{-1}) \otimes \F^{-1} } \paren{ \JM{\F}{\omd} }
&& \text{use P2}
\nonumber \\
& = - \F^{-1} (\x^T\Ft^{-1} \otimes \I) \paren{ \JM{\F}{\omd} }
&& \text{use P3}
\nonumber\\
& = - \F^{-1} \paren{ \JM{\F}{\omd} \z}
&& \text{use P4}
\nonumber\\
& = -(\A'\A + \lambda \T'\T)^{-1} \Big( \eqntrick
\nonumber \\
& \hspace*{5em}
+ \imath \, \diag{\rd} \A' \diag{\A\z} \Big)
&& \text{use \eqref{D,A'A,x,omd}}
\label{e,inv,omd}.
\end{align}


We apply this Jacobian to a vector
by using four NUFFT operations
followed by running CG
to approximate the product
of $\F^{-1}$ times a vector.
Notably, the memory cost of \eqref{e,inv,omd}
is constant w.r.t the number of iterations,
whereas the standard auto-differentiation approach 
has linear memory cost. 
Using the proposed method, 
one may apply enough iterations
to ensure convergence
to a desired tolerance.
This new fast and low-memory Jacobian approximation
is particularly important
for the MRI applications
shown in the following sections.
Without this approximation,
memory cost can be prohibitively large.
\subsection{Sensitivity Maps}
\label{sec,sense}
In multi-coil (parallel) acquisition,
the MRI system model contains another linear operator
\begin{align*}
    \Sm=\bmat{\Sm_{1} \\
    \vdots \\
    \Sm_{\Nc}
    },
\end{align*}
where $\Sm_{i} = \diag{\s_i}$ denotes a diagonal matrix
containing the receiver coil sensitivity map \cite{pruessmann_sense_1999}.
The total number of receiver channels is \Nc.
The system matrix (\Em) for MRI in this case becomes
$(\I_{\Nc} \otimes \A)\Sm$.
Because of the special block-diagonal structure of \Sm,
all the Jacobian matrices in previous sections still hold
by simply replacing \A with \Em.

The Jacobian derivations are as follows.
For the forward operator
(\sref{sec,for}),
one can show
\begin{align*}
\frac{\partial \Em\x}{\partial \omd}
& = \dfrac{\partial \bmat{\A\Sm_{1}\x \\ \vdots \\ \A\Sm_{\Nc}\x}}{\partial \omd}
= \bmat{-\imath \, \diag { \A (\s_{1} \odot \x\odot\rd) } \\
\vdots \\
-\imath \, \diag{\A (\s_{\Nc} \odot \x\odot\rd)}}
\\&
= \imath \, \diag{( \I_{\Nc} \otimes \A )\Sm ( \x\odot\rd )}
\\&
= \imath \, \diag{\Em ( \x\odot\rd )} 
.\end{align*}
The adjoint operator
(\sref{sec,adj})
follows the same proof and produces
\[
\frac{\partial \Em'\y}{\partial \omd} =
\imath \, \diag{\rd} \Em' \diag{\y} 
\label{e,A'y,sense}
.\]
For the gram operator
(\sref{sec,gram})
we have
\begin{align}
  \frac{\partial \Em'\Em\x}{\partial \omd}
& =
\sum_i \dfrac{\partial \Sm_i'\A' \A \Sm_i \x}{\partial \omd}
= \sum_i \Sm_i' \dfrac{\partial \A' \A \Sm_i \x}{\partial \omd}
\nonumber\\&
= \sum_i -\imath \,  \Sm_i'\A' \diag{\A (\Sm \x \odot \rd)}
\nonumber\\&
\quad + \imath \,\Sm_i' \diag{\rd} \A' \diag{\A\Sm_i\x}
\nonumber\\&
= \sum_i -\imath \,  \Sm_i'\A' \diag{\A (\Sm \x \odot \rd)}
\nonumber\\&
\quad + \imath \, \diag{\rd} \Sm_i' \A' \diag{\A\Sm_i\x}
\nonumber\\&
=  -\imath \,  \Em' \diag{\Em (\x \odot \rd)}
\nonumber \\&
\quad + \imath \, \diag{\rd} \Em' \diag{\Em\x} 
\label{e,sens,gram}
.\end{align}

For the inverse of the PSD matrix
(\sref{sec,inv}),
let $\G=\Em'\Em + \lambda \T'\T$ and $\z = \G^{-1}\x$
(in the usual case where the regularizer matrix \T
is designed such that \G is invertible).
Combining \eqref{e,inv,omd} and \eqref{e,sens,gram} produces:
\begin{align*}
  \MoveEqLeft{\dfrac{\partial \paren{ \Em'\Em + \lambda \T'\T }^{-1} \x}{\partial \omd}}
  \nonumber \\
 & = - \G^{-1} \, (\x^T\Gt^{-1} \otimes \I) \, \JM{\G}{\omd} \\
& = -(\Em'\Em + \lambda \T'\T)^{-1} \Big( -\imath \, \Em' \diag{\Em (\z \odot \rd)} \nonumber \\
 & \quad + \imath \, \diag{\rd} \Em' \diag{\Em\z} \Big).
\end{align*}
Again, we apply this Jacobian matrix to a vector
by combining NUFFTs and CG.

\subsection{Field Inhomogeneity}

For MRI scans with long readouts,
one should also consider the effects
of off-resonance
(e.g., $B_0$ field inhomogeneity),
in which case
the system matrix elements are given by
\cite{fessler:10:mbi}
\[
\aij = \Expni{\omi \cdot \rj} \, \Expni{\wj \ti}
,\]
where \wj denotes the field map value at the $j$th voxel
and \ti is the time of the $i$th readout sample.

This form is no longer a Fourier transform operation,
but there are fast and accurate approximations
\cite{fessler:05:tbi}
that enable the use of
$O(N \log N)$ NUFFT steps
and avoid the very slow
$O(N^2)$ matrix-vector multiplication.
Such approximations of system matrix \Em
usually have the form:
\[
\Ef \approx \sum_{l=1}^{L}
\diag{b_{il}} \A(\om) \, \diag{c_{lj}},
\]
where \A denotes the usual (possibly non-uniform) DFT
that is usually
approximated by a NUFFT,
$b_{il} \in \CM$,
and $b_{il} \in \CN$.
It is relatively straightforward
to generalize the Jacobian expressions in this paper
to handle the case of field inhomogeneity, 
by simply replacing $\A$ with $\Ef$,
similar to the sensitivity map case.
\section{Optimizing Sampling Patterns}
\label{s,opt}
For modern MRI systems,
the sampling trajectory \om is a programmable parameter.
Traditionally \om is a geometrical curve controlled by few parameters
(such as radial spokes or spiral leaves),
and its tuning relies on derivative-free optimizers such as grid-search.
In this paper,
we optimize \om
by minimizing a training loss function
from image reconstruction,
where the descent direction of \om
is the negative gradient of that loss \cite{pilot,wang:22:bjork-tmi}.
We adopt such a ``reconstruction loss''
because the terminal goal of
sampling pattern optimization
is to improve image quality.
To learn from large datasets,
the optimization uses
stochastic gradient descent (SGD)-like algorithms.
Additionally,
the loss function may include other terms,
such as a penalty on
the maximum gradient strength 
and slew rate \cite{pilot,wang:22:bjork-tmi}
or peripheral nerve stimulation effects
\cite{wang:22:soo-arxiv}.

For image reconstruction,
consider a convex and smooth regularizer $\R(\cdot)$
for simplicity.
Since the noise statistics are Gaussian,
a typical regularized cost function
used for model-based image reconstruction
is \cite{fessler:10:mbi}
\begin{equation}
\Psi(\x) = \frac{1}{2} \| \A\x - \y \|_2^2 + \R(\x)
.
\label{e,recon}
\end{equation}
During training, the observation \y can be retrospectively simulated
using
$\y = \A(\om)\xtrue$.
For illustration,
consider applying the $k$th step
of gradient descent (GD)
to that cost function:
\begin{align*}
\xkk &= \xk - \alpha \nabla \Psi(\xk)
\\&
= \xk - \alpha \, \A(\om)' \, (\A(\om) \, \xk - \y) - \alpha \nabla \R(\xk)
,\end{align*}
where
$\xk \in \CN$,
$\alpha \in \reals^+$ is the step size.
After $K$ iterations,
we have a reconstructed image (batch)
$\x_K = \x_K(\om) = f(\om, \y)$,
where the reconstruction method $f(\om, \y)$
is a function
of both the data \y
and the sampling pattern \om.
To learn/update the sampling pattern \om,
consider a simple loss function
for a single training example:
\begin{equation}
    L(\om) =  \normii{\x_K(\om) - \xtrue}^2
    \label{e,L1}
\end{equation}
where
\xtrue is the reference fully-sampled image (batch).
Learning \om via backpropagation (or chain-rule)
requires differentiating $L$
w.r.t. the sampling pattern \om,
which in turn
involves Jacobians
of quantities like
$\A(\om)$ that we derived above.

Here we use the forward operator
(\sref{sec,for})
as an example
to illustrate one step in propagation.
As needed in a backpropagation step (Jacobian-vector product, JVP),
the Jacobian \eqref{e,Ax} is multiplied with a gradient vector
$\v = \frac{\partial L}{\partial (\A\x)^*}\in \CM$
calculated in the prior step.
Using \eqref{eqn:wir_chain}, the corresponding JVP is
\begin{equation}
    \frac{\partial L}{\partial \om} = \real{ (-\imath \ \A (\x\odot\rd))' \odot \v }
\label{e,jac,Ax,v}
.\end{equation}
Efficiently computation
can simply apply a NUFFT operation
to $\x \odot \rd$,
followed by a point-wise multiplication with $\v$.
The Gram and PSD inverse (``data consistency")
term in \sref{sec,gram} and \sref{sec,inv}
follow a similar pattern during backpropagation.
See our open-source codes%
\footnote{\url{https://github.com/guanhuaw/Bjork}}
for implementation details.

Although we illustrate the GD algorithm
with a simple smooth regularizer,
more generally,
the reconstruction method
$f(\om,\y)$
can involve
more sophisticated regularizers
such as neural networks
\cite{pilot,wang:22:bjork-tmi} 
or non-smooth sparsity models \cite{lustig:2008:CompressedSensingMRI} used in compressed sensing.
In such cases,
backpropagation uses sub-gradients, instead of gradients,
as is common in stochastic optimization.
The loss JVPs are backpropagated through iterative reconstruction steps
to compute a gradient w.r.t. \om.

The proposed approach
is applicable only to non-Cartesian MRI,
because Cartesian sampling pattern design
is usually a discrete optimization problem,
incompatible with gradient-based methods.
However,
one could optimize
phase-encoding locations continuously
(in 2D or 3D)
with the frequency-encoding direction being fully sampled,
which is a hybrid Cartesian / non-Cartesian approach
\cite{aggarwal:20:jmj}. 


\section{Experiments}
\label{sec:vali}

This section validates the accuracy and efficiency of the proposed methods.
It also showcases the application to MRI sampling trajectory optimization.

\subsection{Accuracy and Efficiency}

The appendix discusses
error bounds
for the Jacobian approximations
of \sref{sec,for}
and \sref{sec,adj}.

We performed numerical experiments
to examine the following test cases:
\[
\frac{\partial \normii{f(\x)}^2}{\partial \omd}
\text{ and }
\frac{\partial \normii{f(\x)}^2}{\partial \x^*},
\]
where $f(\cdot)$ denotes
multiplication
by \A,
by the Gram matrix $\A'\A$,
or by the `inverse of PSD matrix' \blue{(\sref{sec,inv})}
of sensitivity-informed NUFFTs
\blue{(\sref{sec,sense})}.
The Gram and inverse experiments
implicitly test the adjoint operator's approximations.
The $\x$ adopted is
a $40\times40$ patch cropped from the center of
a Shepp–Logan phantom
with random additional phases
uniformly distributed in $[-\pi, \pi]$.
$\Sm$ is a simulated 8-channel sensitivity map,
and \om is one radial spoke
crossing the k-space center.
The Jacobian calculation methods are:
(1) auto-differentiation of NUFFT;
the lookup table operation \cite{dale:01:arl}
is replaced by bilinear interpolation
to enable auto-differentiation,
similar to \cite{pilot},
(2) our approximation described above,
(3) auto-differentiation of exact non-uniform
discrete Fourier transform (NUDFT),
implemented with
single precision.
We regard method 3 (NUDFT) as the ground truth. 
Since NUDFT (in its simplest form)
involves only one exponential function, multiplication and addition
for each element,
its backpropagation introduces minimal numerical errors.
For the PSD inverse,
we applied 20 CG iterations for all three methods,
which was sufficiently close to convergence
based on the residual norm $\|\blmath{r}\|/\|\blmath{b}\|$
(the definition follows \cite[(45)]{shewchuk1994cg}). 

\begin{figure}[hbpt!]
    \centerline{\includegraphics[width=0.8\columnwidth]{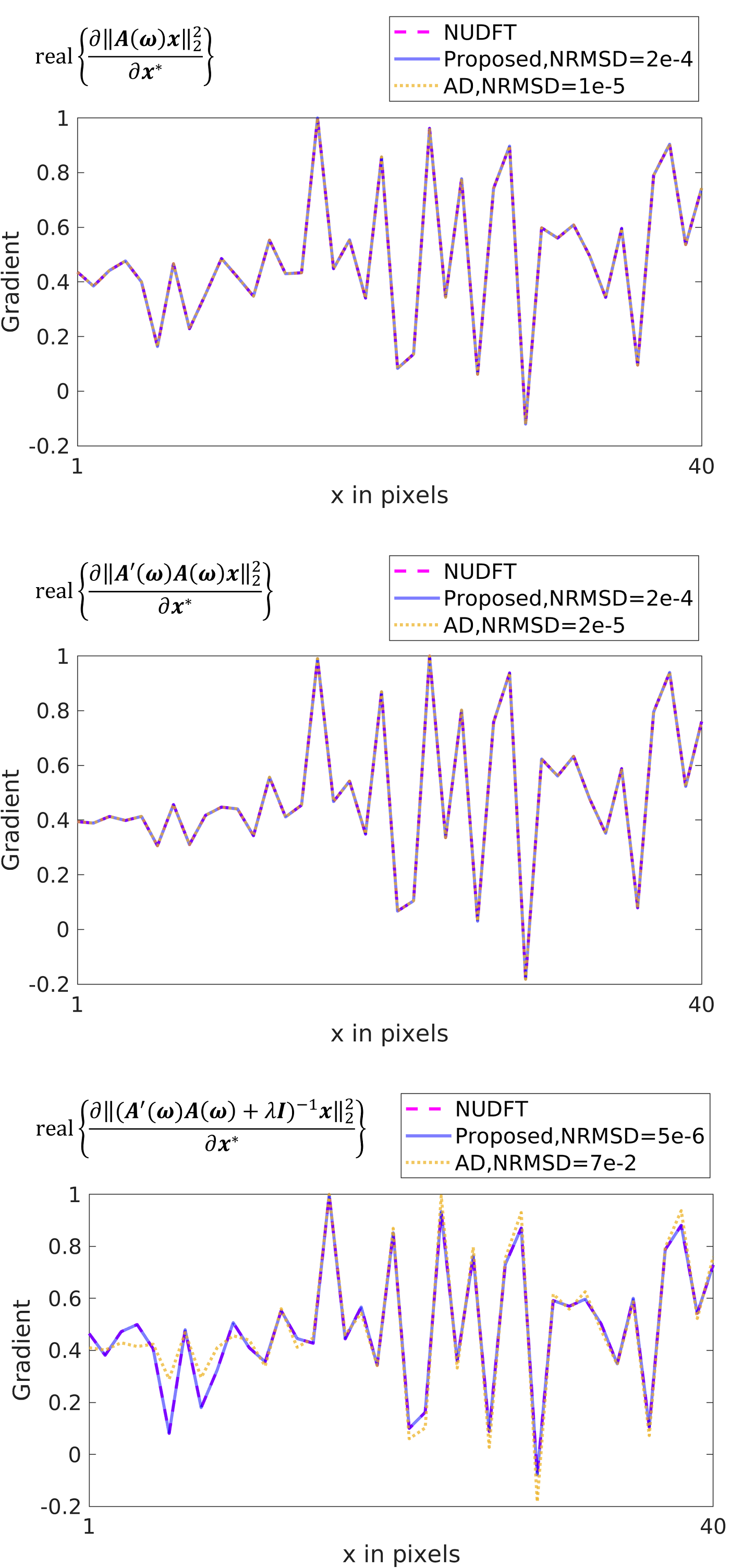}}
    \caption{Examples of gradients w.r.t. $\x^*$ (the real part is plotted).
    Plots show one representative row of a $40 \times 40$ matrix (rescaled to [-1,1]).
    The rows are the forward, Gram, and PSD inverse operator cases.
    The horizontal axis is the pixel index.
    The legend reports
    the normalized root-mean-square difference (NRMSD)
    compared with the reference NUDFT calculation.
    }
    \label{grad_x}
\end{figure}

\begin{figure}[hbpt!]
    \centering
    \includegraphics[width=0.8\columnwidth]{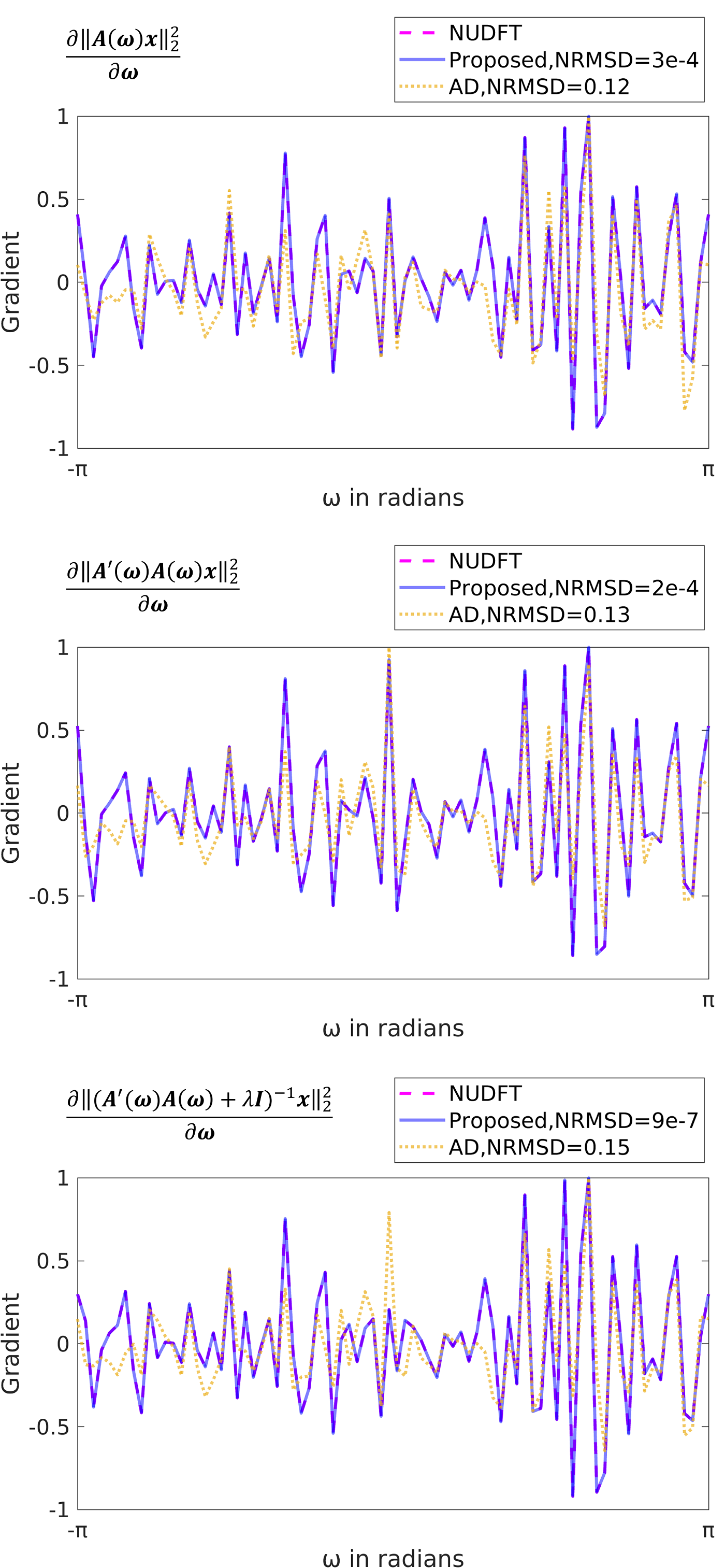}
    \caption{Examples of gradients w.r.t. \om.
    Plots show one spoke of 80 points (rescaled to [-1,1]).
    The rows are the forward, Gram, and PSD inverse operator cases.
    The proposed approximation better matches the gradient of the NUDFT.
    The legend reports
    the normalized root-mean-square difference (NRMSD)
    compared with the reference NUDFT calculation.
    The proposed approach has at least $400\times$ smaller NRMSD
    for this nonlinear case.
    }
    \label{grad_om}
\end{figure}

\fref{grad_x} and \fref{grad_om} illustrate
representative profiles of the gradients
w.r.t. $\x$ and $\om$.
For $\om$,
the auto-differentiation (method~1) approach has larger deviations
from method~3 (NUDFT)
because of the non-differentiability of
interpolation operations
w.r.t. coordinates.
For the gradient w.r.t. $\x$,
both method~1 and method~2
generate accurate results for forward 
and Gram operators.
The reason is that
in method~1 (auto-diff),
the interpolation operation
w.r.t \x is linear,
hence accurately differentiable.
For the PSD inverse, 
method~1 led to a slightly inaccurate gradient,
stemming from the accumulated errors of 
backpropagating CG iterations.

\tref{t,time} and \tref{t,mem}
compare the time
and memory cost 
of methods 1 (auto-diff) and 2 (proposed).
The CPU is Intel(R) Xeon(R) Gold 6138 CPU @ 2.00GHz
and the GPU is an Nvidia(R) RTX2080Ti.
We used PyTorch 1.9.1 and torchkbnufft 1.1.0.
The memory usage was tracked by $\texttt{torch.cuda.max\_memory\_allocated}$
on Nvidia GPUs.
We implemented the numerical experiments with
torchkbnufft\footnote{\url{https://github.com/mmuckley/torchkbnufft}} \cite{muckley:20:tah} and MIRTorch\footnote{\url{https://github.com/guanhuaw/MIRTorch}} toolboxes.

Our method is much faster than auto-differentiation
on both GPUs and CPUs,
and uses less memory.
Importantly,
the PSD inverse Jacobian
is impractical for the 3D case,
whereas the proposed approach
fit comfortably in GPU's onboard memory.

\begin{table}[htbp]
\caption{Computation time of the test case.}
\label{t,time}
\centering
\begin{tabular}{l|rr|rr}
\hline
& \multicolumn{2}{c|}{Gram}   &  \multicolumn{2}{|c}{Inverse}  \\\cline{2-5}
& auto-diff & proposed  & auto-diff & proposed  \\ \hline
Large image - GPU  &     0.3s      &    \textbf{0.2s}  &  4.3s & \textbf{2.5s}\\
Small image - GPU  &      0.1s     &    \textbf{0.1s}    & 1.3s & \textbf{0.9s}     \\
Large image - CPU  &    5.2s       &    \textbf{1.7s}      &276.2s & \textbf{48.5s}     \\
Small image - CPU  &   0.8s       &    \textbf{0.5s}     &27.4s & \textbf{6.9s}      \\ \hline
\multicolumn{3}{l}{Large size: $400\times400$; small size: $40\times40$}\\
\multicolumn{5}{l}{
20 CG iterations were applied in the PSD inverse cases.}\\
\end{tabular}
\end{table}

\begin{table}[htbp]
\caption{Memory use of the test case.}
\label{t,mem}
\centering
\begin{tabular}{l|rr|rr}
\hline
& \multicolumn{2}{c|}{Gram}   &  \multicolumn{2}{|c}{Inverse}  \\\cline{2-5}
& standard & proposed  & standard & proposed  \\ \hline
Small  &    3.1MB     &    \textbf{2.8MB}    & 145.7MB & \textbf{2.9MB} \\ 
Large  &    375.9MB      &    \textbf{267.5MB}  & 5673.2MB & \textbf{272.0MB}\\
3D  &    N/A      &    \textbf{10.1GB}  & N/A & \textbf{10.8GB}\\ \hline
\multicolumn{5}{l}{Large size: $400\times400$; small size: $40\times40$;}
\\
\multicolumn{5}{l}{3D size: $200\times200\times100$.}
\\
\multicolumn{5}{l}{
N/A: the memory usage was too large for a single GPU.
}\\
\multicolumn{5}{l}{
20 CG iterations were applied in the PSD inverse cases.
}\\
\end{tabular}
\end{table}

\subsection{MRI Trajectory Optimization}
\label{sebsec:mriopt}
This experiment optimized the MRI sampling trajectory using the
proposed Jacobian approximations
and stochastic optimization.
The reconstruction methods \eqref{e,recon}
here consider two types of algorithms,
namely smooth (regularized) least-squares
reconstruction
and sparsity-based reconstruction.

The smooth reconstruction method uses the 
cost function
\[
\Psi(\x) = \frac{1}{2} \| \Em(\om) \x - \y \|_2^2 +\frac{\lambda}{2} \| \T \x \|_2^2,
\]
where $\T$ is a finite-difference operator 
encouraging smoothness.
Correspondingly, the reconstructed image is:
\[
\x_K = (\Em'\Em + \lambda \T'\T)^{-1} \Em'\y,
\]
which we solved using CG.
The following sections refer to this method as 
quadratic penalized least-squares (QPLS).
We also implemented a 
simpler case,
where $\T = \I$,
which is referred as CG-SENSE \cite{maier:2021:CGSENSERevisitedResults}.
In both scenarios,
we set
$\lambda$ to $10^{-3}$ empirically
and still applied 20 CG iterations.
The initialization of CG
used the density compensated reconstruction \cite{hoge1997density}.

The sparsity-based compressed (CS) sensing algorithm
adopts a wavelets-based sparsity penalty,
and has the following objective function
\[
\Psi(\x) = \frac{1}{2} \| \Em(\om) \x - \y \|_2^2 +\lambda \| \W \x \|_1,
\]
where \W is an orthogonal DWT matrix
and we set $\lambda = 10^{-5}$ empirically.
We used 40 iterations
of the proximal optimized gradient method (POGM)
\cite{kim:2018:AdaptiveRestartOptimized,fessler:20:omf}
to solve this non-smooth optimization problem.

For the purpose
of comparing trajectories and image quality,
we also applied
the proposed approximations
to an unrolled neural network (UNN)
reconstruction method 
that followed the definition 
of \cite{aggarwal:19:mmb} (\cite{wang:22:bjork-tmi} extends
it to non-Cartesian cases.)
We used the same network configuration as in \cite{wang:22:bjork-tmi}.

To optimize the k-space trajectory
for each of these reconstruction methods,
the training loss \eqref{e,L1} is:
\begin{align*}
L(\om)  = & \normii{\x_K(\om) - \xtrue}^2 \\
&+ \mu_1 \blmath{\phi}_{\gamma \Delta t \gmax}(|\D_1\om|) 
+ \mu_2 \blmath{\phi}_{\gamma\Delta t^{2} \smax}(|\D_2\om|),
\end{align*}
where \xtrue is the conjugate phase reconstruction
of fully sampled Cartesian data \cite{noll:2005:cpm}.
The second and third terms applied a soft constraint on
gradient strength
and slew rate according to \cite[Eqn.~2]{wang:22:bjork-tmi},
where $\blmath{\phi}_\lambda(|\x|)
= \sum \max(|\x|-\lambda, 0)$.
The maximum gradient strength (\gmax) was 5 Gauss/cm 
and the maximum slew rate (\smax) was 15 Gauss/cm/ms.
$\mu_1 = \mu_2 = 10$.
We estimated sensitivity maps in \Em using ESPIRiT \cite{espirit},
and simulated noiseless raw signals $\y = \Em(\om) \xtrue$
retrospectively w.r.t. 
candidate trajectories.
The training used the fastMRI brain dataset \cite{fastmri} containing 
15902 T1w slices, 16020 T2w slices, and 3311 FLAIR slices
cropped to size $320 \times 320$.
The number of coils ranges from 2 to 28. 
We used the Adam optimizer
\cite{kingma:2017:AdamMethodStochastic},
with step size $10^{-4}$
and mini-batch size $12$.
We used 6 epochs for training model-based methods (CG-SENSE, QPLS and CS)
and 60 epochs for the UNN training.
The initialization of learned trajectories
was an under-sampled radial trajectory 
in all experiments.
The initialization had 16 ``spokes''
and each spoke was 5ms long with 1280 sampling points.
We also adopted the k-space parameterization trick 
detailed in \cite[Eqn.~3]{wang:22:bjork-tmi}
to avoid sub-optimal local minima,
and parameterized each shot with 40 quadratic spline kernels.

\begin{figure*}[hbpt!]
    \centering
    \includegraphics[width=0.95\textwidth]{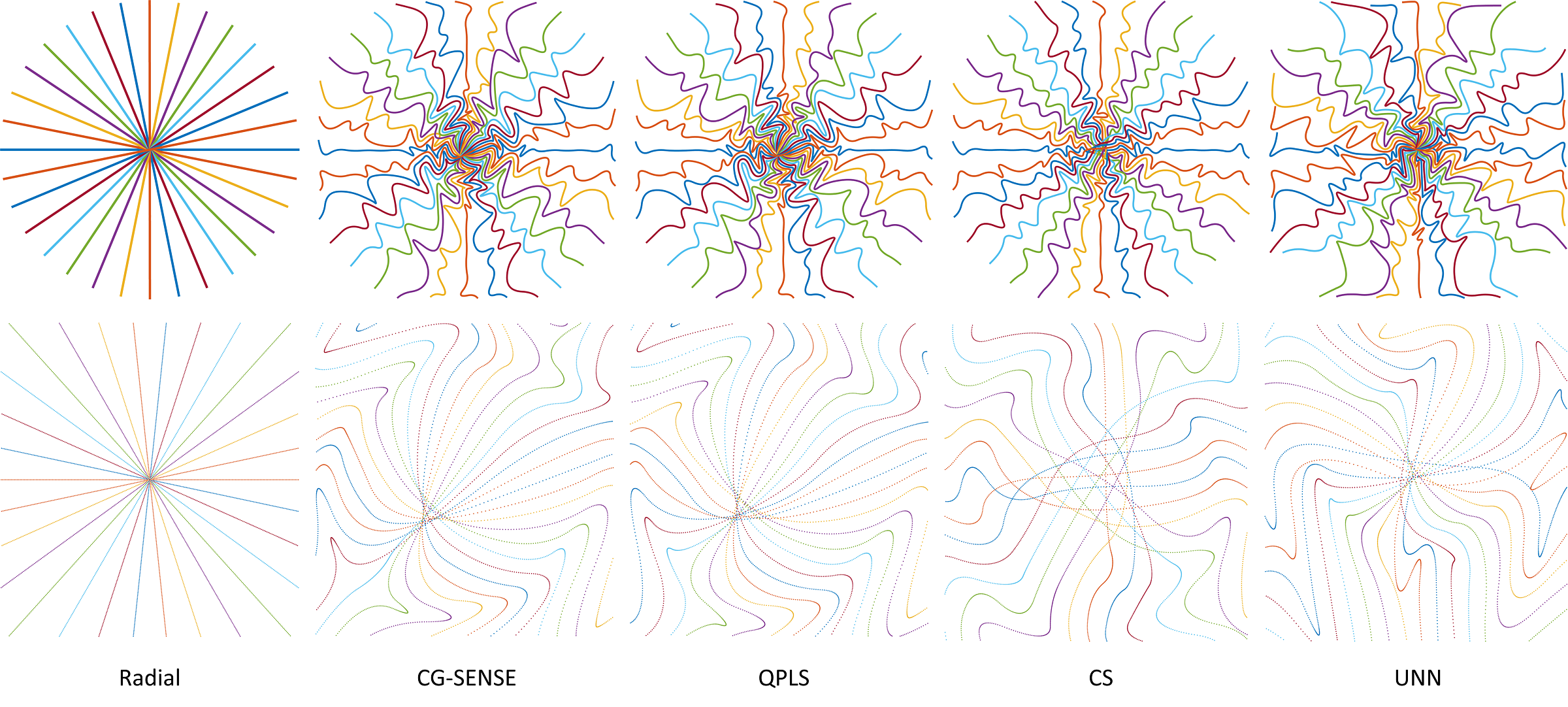}
    \caption{Optimized sampling trajectories for several iterative reconstruction methods.
    The left column shows the uniform radial initialization.
    The second row shows the $8\times$ zoomed-in central k-space.}
    \label{grad_recon}
\end{figure*}

\begin{figure*}[hbpt!]
    \centering
    \includegraphics[width=0.95\textwidth]{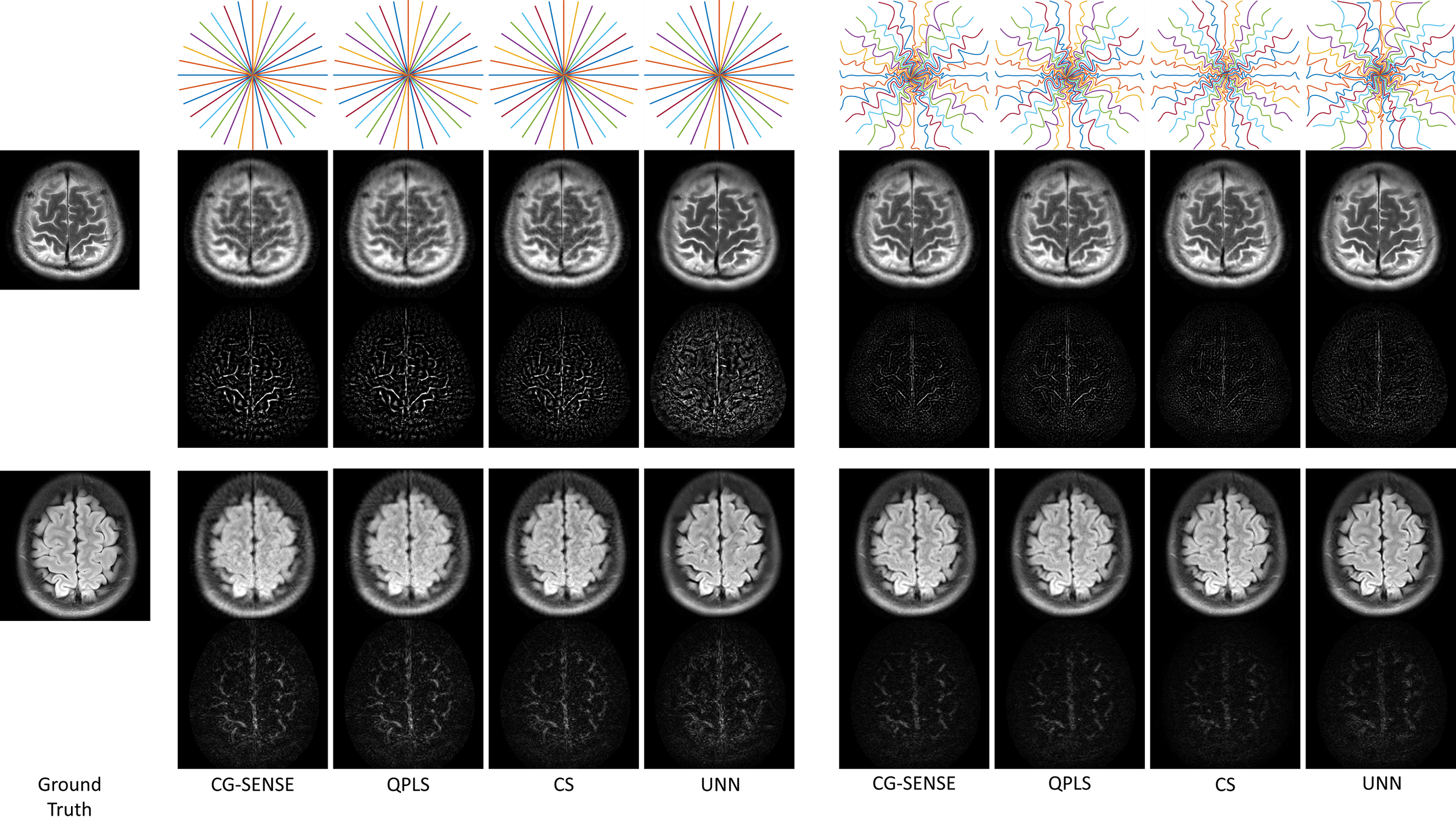}
    \caption{Examples of the reconstructed images with unoptimized (left) and optimized trajectories (right).
    Rows 3 and 5 show corresponding error maps.}
    \label{grad_sample}
\end{figure*}

\fref{grad_recon} showcases the trajectories optimized
for each of the reconstruction methods.
The centers of trajectories optimized with quadratic
regularizers (CG-SENSE and QPLS) are not aligned with
the k-space origin.
We hypothesize that regularizers (and corresponding
iterative algorithms) handle image phases differently,
resulting in distinct trajectory centers.

\tref{tab:quan} reports the average image reconstruction quality
(PSNR and SSIM \cite{hore:2010:ImageQualityMetrics},
fully sampled image as the ground truth) on 500 test slices.
It also showcases the image quality
of these learned trajectories
with reconstruction methods different from the training phase.
All learned trajectories led to improved reconstruction quality
compared to the initial radial trajectory (unopt.),
even with different reconstruction methods.
Importantly, the same reconstruction algorithm across training
and test led to the greatest improvement (the bold diagonal entries).
\fref{grad_sample} shows reconstruction examples.

\begin{table}[htbp!]
\caption{Average reconstruction quality on test set
with trajectories learned for different reconstruction methods.}
\begin{tabular}{llllll}
\multicolumn{6}{l}{SSIM}\\
\hline
\backslashbox{Test}{Learn} & QPLS & SENSE & CS & UNN & unopt.\\ \hline
QPLS   &\textbf{0.963}     & 0.963    & 0.962  &0.961 & 0.947  \\
SENSE   &0.964     & \textbf{0.964}    &  0.963 & 0.961 &  0.946 \\
CS &0.962     & 0.963    &\textbf{0.966} &  0.964  &  0.946 \\
UNN & 0.960 & 0.960 &0.958 &\textbf{0.964} & 0.950\\
\hline
\\
\multicolumn{6}{l}{PSNR (in dB)}\\
\hline
\backslashbox{Test}{Learn} & QPLS & SENSE & CS & UNN & unopt.\\ \hline
QPLS   &    \textbf{35.1} &   35.1  &  34.9 & 35.0 & 33.1   \\
SENSE   & 35.2    & \textbf{35.2}    & 34.9 &  35.1& 33.1  \\
CS & 34.8     &34.9     &  \textbf{35.4} &35.2 & 33.0  \\
UNN &34.6 &34.6 &34.5 & \textbf{35.0} &33.5 \\
\hline
\multicolumn{6}{l}{Learn: the reconstruction method
the trajectory is jointly trained with.}\\
\multicolumn{6}{l}{Test: the reconstruction method in the test phase.}\\
\multicolumn{6}{l}{unopt.: the undersampled radial trajectory (unoptimized initialization).}
\end{tabular}
\label{tab:quan}
\end{table}

\begin{figure}[hbpt!]
    \centerline{\includegraphics[width=0.9\columnwidth]{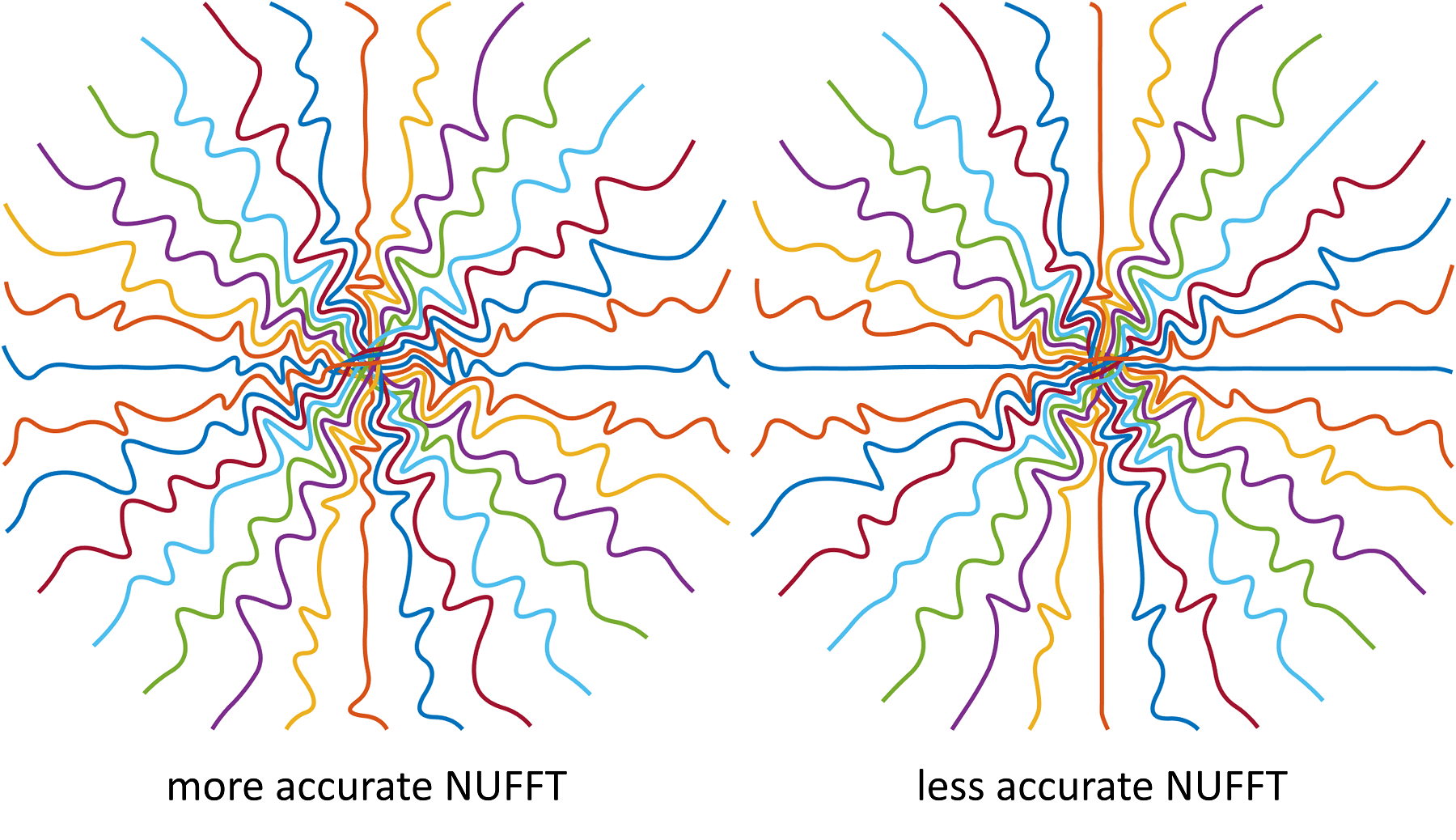}}
    \caption{Learned trajectories with different NUFFT accuracies.}
    \label{fig:accu}
\end{figure}

\subsection{Accelerated Learning with Low-Accuracy NUFFT}

The major computation cost of
trajectory learning is proportional to NUFFTs
and their Jacobian calculations.
An empirical acceleration method is to use faster NUFFT approximations (low over-sampling factors
and/or small interpolation neighborhoods)
in training.
Later,
when the learned trajectory is deployed on test data
or prospectively acquired data,
one could use default NUFFT accuracy.
We investigated learning trajectories 
with two different NUFFT accuracies:
(1) gridding size = $1.25\times$ image size,
interpolation kernel size = 5 and
(2) gridding size = $2\times$ image size,
interpolation kernel size = 6
which is a commonly used setting.
On our GPUs,
the lower-accuracy setting was $1.4\times$ faster than
the higher-accuracy one.
We used the CS-based reconstruction
and corresponding training strategy
described in the previous subsection.
\fref{fig:accu} shows the trajectory
optimized for the two NUFFT accuracy levels.
To compare the trajectory optimized 
by these two settings,
we used
the reconstruction image quality
as the evaluation metric.
We simulated and reconstructed images using 
the two trajectories
on the test data 
(same as the previous experiment).
The trajectories optimized
with the ``low accuracy'' and ``high accuracy'' NUFFT
had mean PSNR values
of 35.4$\pm$4.6 dB and 35.4$\pm$4.7 dB.

\section{Discussion}
\label{sec:disc}
This paper presents a model-based approximation
of Jacobian matrices involving NUFFTs.
Compared to direct auto-differentiation,
the proposed method is faster,
needs less memory,
and better approximates the reference NUDFT results.
As discussed in \ref{sec,model},
the error of auto-differentiation is not a software limitation,
but rather a problem
that stems from the non-differentiability
of interpolation or lookup table operations.
NUFFT alternatives such as re-gridding
or filtered back-projection
also suffer from similar non-differentiability issues
and are less effective than (NUFFT-based) iterative reconstruction.
Our previous studies \cite[Fig.~14]{wang:22:bjork-tmi} compared 
the trajectory optimization results of
the proposed method
and standard auto-differentiation.
The trajectory optimized by the proposed approximation
generated superior image quality,
and conformed better to the empirical criteria \cite{sparklingmrm, boyer2016generation}:
sampling points should not overlap
or be too distant from each other.

Sampling patterns learned with different reconstruction methods
showed distinct characteristics in \sref{sebsec:mriopt}.
This phenomenon was also observed in 
previous literature \cite{zibetti2020fast,gozcu:2019:RSP}.
The differences in sampling patterns may
stem from different regularizers,
as well as different iterative algorithms.
Importantly, as shown in \tref{tab:quan},
synergistic sampling and reconstruction
led to the best image quality.
Several previous studies \cite{pilot,wang:22:bjork-tmi,aggarwal:20:jmj}
only used NN-based reconstruction methods,
while the stability and generalizability
of NN-based reconstruction are still
being investigated.
In comparison,
using our method delineated in \sref{s,opt},
one may optimize trajectories
for model-based reconstruction methods
that may be more robust.
Our results show that
with a suitably tailored sampling pattern,
traditional model-based reconstruction
can compete with NN-based reconstruction,
reinforcing related observations
in recent studies
\cite{gu:22:rlw}.
Additionally, sampling optimization
for model-based reconstruction 
requires less training data
than for NN-based reconstruction.
This property is beneficial for medical imaging
where the data availability is often limited.

The training used discrete-space image datasets,
whereas the actual objects in practice are continuous.
Ideally, using an accurate continuous image model
could better approximate the actual situation.
This implicit bias is common for learning-based methods,
and may lead to suboptimal results,
such as the backtracking in the edge/corner of k-space (\fref{grad_recon}).
The training also ignored physical processes such as 
relaxation and magnetization transfer. 
Future studies may consider these processes
in the forward system model.
The mismatch or domain shift from training 
to prospective scans may influence the results.
For example, there exist differences in protocols
(RF pulses, FOVs, and resolutions),
hardware (field strengths and Tx/Rx coils),
system imperfections (eddy currents, gradient non-linearity, and inhomogeneity),
demography, and pathology.
Our previous studies \cite{wang:22:bjork-tmi} tested
the optimized trajectory in a prospective 
in-vivo experiment,
and discussed practical issues, 
including eddy currents,
and contrast/SNR mismatch between the training set
and prospective protocols.
Subsequent studies should evaluate
the robustness of learned sampling trajectories in more scenarios.

\section*{Acknowledgment}
The authors gratefully thank Dr.~Douglas Noll, Dr.~Tianrui Luo, Naveen Murthy, Yuran Zhu,
and the anonymous reviewers for helpful advice.

\bibliographystyle{IEEEtran}
\bibliography{refs}

\begin{thebibliography}{10}
\providecommand{\url}[1]{#1}
\csname url@samestyle\endcsname
\providecommand{\newblock}{\relax}
\providecommand{\bibinfo}[2]{#2}
\providecommand{\BIBentrySTDinterwordspacing}{\spaceskip=0pt\relax}
\providecommand{\BIBentryALTinterwordstretchfactor}{4}
\providecommand{\BIBentryALTinterwordspacing}{\spaceskip=\fontdimen2\font plus
\BIBentryALTinterwordstretchfactor\fontdimen3\font minus
  \fontdimen4\font\relax}
\providecommand{\BIBforeignlanguage}[2]{{%
\expandafter\ifx\csname l@#1\endcsname\relax
\typeout{** WARNING: IEEEtran.bst: No hyphenation pattern has been}%
\typeout{** loaded for the language `#1'. Using the pattern for}%
\typeout{** the default language instead.}%
\else
\language=\csname l@#1\endcsname
\fi
#2}}
\providecommand{\BIBdecl}{\relax}
\BIBdecl

\bibitem{munson:84:irf}
D.~C. Munson and J.~L. Sanz, ``Image reconstruction from frequency-offset
  {Fourier} data,'' \emph{{Proc. IEEE}}, vol.~72, no.~6, pp. {661--9}, Jun.
  1984.

\bibitem{bronstein:02:rid}
M.~M. Bronstein, A.~M. Bronstein, M.~Zibulevsky, and H.~Azhari,
  ``Reconstruction in diffraction ultrasound tomography using nonuniform
  {FFT},'' \emph{{IEEE Trans. Med. Imag.}}, vol.~21, no.~11, pp. {1395--1401},
  Nov. 2002.

\bibitem{matej:04:iti}
S.~Matej, J.~A. Fessler, and I.~G. Kazantsev, ``Iterative tomographic image
  reconstruction using {Fourier-based} forward and back- projectors,''
  \emph{{IEEE Trans. Med. Imag.}}, vol.~23, no.~4, pp. {401--12}, Apr. 2004.

\bibitem{wright:14:ncr}
K.~L. Wright, J.~I. Hamilton, M.~A. Griswold, V.~Gulani, and N.~Seiberlich,
  ``{Non-Cartesian parallel imaging reconstruction},'' \emph{J. Magn. Reson.
  Imag.}, vol.~40, no.~5, pp. 1022--1040, 2014.

\bibitem{fessler:10:mbi}
J.~A. Fessler, ``Model-based image reconstruction for {MRI},'' \emph{{IEEE Sig.
  Proc. Mag.}}, vol.~27, no.~4, pp. {81--9}, Jul. 2010.

\bibitem{fessler:03:nff}
J.~A. Fessler and B.~P. Sutton, ``Nonuniform fast {Fourier} transforms using
  min-fmax interpolation,'' \emph{{IEEE Trans. Sig. Proc.}}, vol.~51, no.~2,
  pp. {560--74}, Feb. 2003.

\bibitem{yang:14:mso}
Z.~Yang and M.~Jacob, ``Mean square optimal {NUFFT} approximation for efficient
  {non-Cartesian} {MRI} reconstruction,'' \emph{{J. Mag. Res.}}, vol. 242, pp.
  {126--35}, May 2014.

\bibitem{zibetti2020fast}
M.~V.~W. Zibetti, G.~T. Herman, and R.~R. Regatte, ``Fast data-driven learning
  of parallel {{MRI}} sampling patterns for large scale problems,'' \emph{Sci
  Rep}, vol.~11, no.~1, p. 19312, Sep. 2021.

\bibitem{sanchez:2020:ScalableLearningBasedSampling}
T.~Sanchez \emph{et~al.}, ``Scalable {{learning}}-{{based sampling
  optimization}} for {{compressive dynamic MRI}},'' in \emph{2020 {{IEEE Intl.
  Conf.}} on {{Acous.}}, {{Speech}} and {{Sig. Proc.}} ({{ICASSP}})}, May 2020,
  pp. 8584--8588.

\bibitem{gozcu:2019:RSP}
B.~G{\"o}zc{\"u}, T.~Sanchez, and V.~Cevher, ``Rethinking {{Sampling}} in
  {{Parallel MRI}}: {{A Data-Driven Approach}},'' in \emph{2019 27th {{Euro.
  Sig. Proc. Conf.}} ({{EUSIPCO}})}, Sep. 2019, pp. 1--5.

\bibitem{cao:1993:Feature}
Y.~Cao and D.~N. Levin, ``Feature-recognizing {MRI},'' \emph{Magn. Reson.
  Med.}, vol.~30, no.~3, pp. 305--317, 1993.

\bibitem{knoll:11:ars}
F.~Knoll, C.~Clason, C.~Diwoky, and R.~Stollberger, ``Adapted random sampling
  patterns for accelerated {MRI},'' \emph{{Mag. Res. Mat. Phys. Bio. Med.}},
  vol.~24, no.~1, pp. {43--50}, Feb. 2011.

\bibitem{bahadir:2020:LOUPE}
C.~D. Bahadir, A.~Q. Wang, A.~V. Dalca, and M.~R. Sabuncu,
  ``Deep-{{learning}}-{{based optimization}} of the {{under}}-{{sampling
  pattern}} in {{MRI}},'' \emph{IEEE Trans. Comput. Imag.}, vol.~6, pp.
  1139--1152, 2020.

\bibitem{sherry:20:lts}
F.~Sherry, M.~Benning, J.~C. D.~. Reyes, M.~J. Graves, G.~Maierhofer,
  G.~Williams, C.-B. {Schonlieb}, and M.~J. Ehrhardt, ``Learning the sampling
  pattern for {MRI},'' \emph{{IEEE Trans. Med. Imag.}}, vol.~39, no.~12, pp.
  {4310--21}, Dec. 2020.

\bibitem{huijben:2020:LearningSamplingModelBased}
I.~A.~M. Huijben, B.~S. Veeling, and R.~J.~G. van Sloun, ``Learning sampling
  and model-based signal recovery for compressed sensing {MRI},'' in \emph{2020
  {{IEEE Intl. Conf.}} on {{Acous.}}, {{Speech}} and {{Sig. Proc.}}
  ({{ICASSP}})}, May 2020, pp. 8906--8910.

\bibitem{seeger:2010:OptimizationKspaceTrajectories}
M.~Seeger, H.~Nickisch, R.~Pohmann, and B.~Sch{\"o}lkopf, ``Optimization of
  {k-space} trajectories for compressed sensing by {{Bayesian}} experimental
  design,'' \emph{Magn. Reson. Med.}, vol.~63, no.~1, pp. 116--126, 2010.

\bibitem{haldar:2019:OEDIPUS}
J.~P. Haldar and D.~Kim, ``{{OEDIPUS}}: {{An Experiment Design Framework}} for
  {{Sparsity-Constrained MRI}},'' \emph{IEEE Trans. Med. Imaging}, vol.~38,
  no.~7, pp. 1545--1558, Jul. 2019.

\bibitem{vonkienlin:1991:Spectral}
M.~{von Kienlin} and R.~Mejia, ``Spectral localization with optimal pointspread
  function,'' \emph{J. Magn. Reson. (1969)}, vol.~94, no.~2, pp. 268--287, Sep.
  1991.

\bibitem{gao:2000:OptimalKspaceSampling}
Y.~Gao and S.~Reeves, ``Optimal k-space sampling in {{MRSI}} for images with a
  limited region of support,'' \emph{IEEE Trans. Med. Imag.}, vol.~19, no.~12,
  pp. 1168--1178, Dec. 2000.

\bibitem{xu2005optimal}
D.~Xu, M.~Jacob, and Z.~Liang, ``Optimal sampling of k-space with cartesian
  grids for parallel {MR} imaging,'' in \emph{{Proc. Intl. Soc. Mag. Res.
  Med.}}, 2005, p. 2450.

\bibitem{levine2017fly}
E.~Levine and B.~Hargreaves, ``On-the-fly adaptive {k-Space} sampling for
  linear {MRI} reconstruction using moment-based spectral analysis,''
  \emph{IEEE Trans. Med. Imag.}, vol.~37, no.~2, pp. 557--567, 2017.

\bibitem{pilot}
T.~Weiss, O.~Senouf, S.~Vedula, O.~Michailovich, M.~Zibulevsky, and
  A.~Bronstein, ``{PILOT}: Physics-informed learned optimized trajectories for
  accelerated {MRI},'' \emph{MELBA}, pp. 1--23, 2021.

\bibitem{aggarwal:20:jmj}
H.~K. Aggarwal and M.~Jacob, ``{J-MoDL:} {Joint} model-based deep learning for
  optimized sampling and reconstruction,'' \emph{{IEEE J. Sel. Top. Sig.
  Proc.}}, vol.~14, no.~6, pp. {1151--62}, Oct. 2020.

\bibitem{wang:22:bjork-tmi}
G.~Wang, T.~Luo, J.-F. Nielsen, D.~C. Noll, and J.~A. Fessler, ``B-{{Spline
  Parameterized Joint Optimization}} of {{Reconstruction}} and {{K-Space
  Trajectories}} ({{BJORK}}) for {{Accelerated 2D MRI}},'' \emph{IEEE Trans.
  Med. Imag.}, vol.~41, no.~9, pp. 2318--2330, Sep. 2022.

\bibitem{scope2022efficient}
E.~Scope~Crafts, H.~Lu, H.~Ye, L.~L. Wald, and B.~Zhao, ``An efficient approach
  to optimal experimental design for magnetic resonance fingerprinting with
  {B-splines},'' \emph{Magn. Reson. Med.}, vol.~88, no.~1, pp. 239--253, 2022.

\bibitem{jordan2021automated}
S.~P. Jordan, S.~Hu, I.~Rozada, D.~F. McGivney, R.~Boyacio{\u{g}}lu, D.~C.
  Jacob, S.~Huang, M.~Beverland, H.~G. Katzgraber, M.~Troyer \emph{et~al.},
  ``Automated design of pulse sequences for magnetic resonance fingerprinting
  using physics-inspired optimization,'' \emph{Proc. Natl. Acad. Sci}, vol.
  118, no.~40, p. e2020516118, 2021.

\bibitem{dale:01:arl}
B.~Dale, M.~Wendt, and J.~L. Duerk, ``A rapid look-up table method for
  reconstructing {MR} images from arbitrary {K-space} trajectories,''
  \emph{{IEEE Trans. Med. Imag.}}, vol.~20, no.~3, pp. {207--17}, Mar. 2001.

\bibitem{beatty:05:rgr}
P.~J. Beatty, D.~G. Nishimura, and J.~M. Pauly, ``Rapid gridding reconstruction
  with a minimal oversampling ratio,'' \emph{{IEEE Trans. Med. Imag.}},
  vol.~24, no.~6, pp. {799--808}, Jun. 2005.

\bibitem{pruessmann_sense_1999}
K.~P. Pruessmann, M.~Weiger, M.~B. Scheidegger, and P.~Boesiger, ``{SENSE}:
  sensitivity encoding for fast {MRI},'' \emph{Magn. Reson. Med.}, vol.~42,
  no.~5, pp. 952--962, 1999.

\bibitem{sutton:2003:FastIterativeImagea}
B.~Sutton, D.~Noll, and J.~Fessler, ``Fast, iterative image reconstruction for
  {{MRI}} in the presence of field inhomogeneities,'' \emph{IEEE Trans. Med.
  Imaging}, vol.~22, no.~2, pp. 178--188, Feb. 2003.

\bibitem{zehni:20:jar}
M.~Zehni, L.~Donati, E.~Soubies, Z.~Zhao, and M.~Unser, ``Joint angular
  refinement and reconstruction for single-particle {cryo-EM},'' \emph{{IEEE
  Trans. Im. Proc.}}, vol.~29, pp. {6151--63}, 2020.

\bibitem{wang:22:rmb}
\BIBentryALTinterwordspacing
G.~Wang, D.~C. Noll, and J.~A. Fessler, ``Reconstruction may benefit from
  tailored sampling trajectories: optimizing {non-Cartesian} trajectories for
  model-based reconstruction,'' in \emph{{Proc. Intl. Soc. Mag. Res. Med.}},
  2022, p. 5011. [Online]. Available:
  \url{https://submissions.mirasmart.com/ISMRM2022/Itinerary/Files/PDFFiles/5011.html}
\BIBentrySTDinterwordspacing

\bibitem{remmert1991theory}
R.~Remmert, \emph{Theory of complex functions}.\hskip 1em plus 0.5em minus
  0.4em\relax Springer Science \& Business Media, 1991, vol. 122.

\bibitem{kreutz2009complex}
K.~Kreutz-Delgado, ``The complex gradient operator and the {CR}-calculus,''
  \emph{arXiv preprint arXiv:0906.4835}, 2009.

\bibitem{hjorungnes2007complex}
A.~Hjorungnes and D.~Gesbert, ``Complex-valued matrix differentiation:
  Techniques and key results,'' \emph{IEEE Trans. Sig. Proc.}, vol.~55, no.~6,
  pp. 2740--2746, 2007.

\bibitem{wiki:numerator}
\BIBentryALTinterwordspacing
{Wikipedia contributors}, ``Numerator layout notation,'' 2022, [Accessed
  2022-11-08]. [Online]. Available:
  \url{https://en.wikipedia.org/wiki/Matrix_calculus\#Numerator-layout_notation}
\BIBentrySTDinterwordspacing

\bibitem{magnus2019matrix}
J.~R. Magnus and H.~Neudecker, \emph{Matrix differential calculus with
  applications in statistics and econometrics}.\hskip 1em plus 0.5em minus
  0.4em\relax John Wiley \& Sons, 2019.

\bibitem{hestenes:1969:MultiplierGradientMethodsa}
M.~R. Hestenes, ``Multiplier and gradient methods,'' \emph{J. Optim. Theory
  Appl.}, vol.~4, no.~5, pp. 303--320, Nov. 1969.

\bibitem{aggarwal:19:mmb}
H.~K. Aggarwal, M.~P. Mani, and M.~Jacob, ``{MoDL:} model-based deep learning
  architecture for inverse problems,'' \emph{{IEEE Trans. Med. Imag.}},
  vol.~38, no.~2, pp. {394--405}, Feb. 2019.

\bibitem{ramani:11:pmi}
S.~Ramani and J.~A. Fessler, ``Parallel {MR} image reconstruction using
  augmented {Lagrangian} methods,'' \emph{{IEEE Trans. Med. Imag.}}, vol.~30,
  no.~3, pp. {694--706}, Mar. 2011.

\bibitem{chan:17:pap}
S.~H. Chan, X.~Wang, and O.~A. Elgendy, ``Plug-and-play {ADMM} for image
  restoration: fixed-point convergence and applications,'' \emph{{IEEE Trans.
  Computational Imaging}}, vol.~3, no.~1, pp. {84--98}, Mar. 2017.

\bibitem{fessler:05:tbi}
J.~A. Fessler, S.~Lee, V.~T. Olafsson, H.~R. Shi, and D.~C. Noll,
  ``Toeplitz-based iterative image reconstruction for {MRI} with correction for
  magnetic field inhomogeneity,'' \emph{{IEEE Trans. Sig. Proc.}}, vol.~53,
  no.~9, pp. {3393--402}, Sep. 2005.

\bibitem{wang:22:soo-arxiv}
\BIBentryALTinterwordspacing
G.~Wang, J.-F. Nielsen, J.~A. Fessler, and D.~C. Noll, ``Stochastic
  optimization of {3D} {non-Cartesian} sampling trajectory {(SNOPY)},'' 2022.
  [Online]. Available: \url{http://arxiv.org/abs/2209.11030}
\BIBentrySTDinterwordspacing

\bibitem{lustig:2008:CompressedSensingMRI}
M.~Lustig, D.~L. Donoho, J.~M. Santos, and J.~M. Pauly, ``Compressed sensing
  {MRI},'' \emph{IEEE Signal Process. Mag.}, vol.~25, no.~2, pp. 72--82, Mar.
  2008.

\bibitem{shewchuk1994cg}
\BIBentryALTinterwordspacing
J.~R. Shewchuk \emph{et~al.}, ``An introduction to the conjugate gradient
  method without the agonizing pain,'' 1994. [Online]. Available:
  \url{https://www.cs.cmu.edu/~quake-papers/painless-conjugate-gradient.pdf}
\BIBentrySTDinterwordspacing

\bibitem{muckley:20:tah}
M.~J. Muckley, R.~Stern, T.~Murrell, and F.~Knoll, ``{TorchKbNufft}: A
  high-level, hardware-agnostic non-uniform fast fourier transform,'' in
  \emph{ISMRM Workshop on Data Sampling \& Image Reconstruction}, 2020.

\bibitem{maier:2021:CGSENSERevisitedResults}
O.~Maier \emph{et~al.}, ``{{CG}}-{{SENSE}} revisited: {{Results}} from the
  first {{ISMRM}} reproducibility challenge,'' \emph{Magn. Reson. Med.},
  vol.~85, no.~4, pp. 1821--1839, 2021.

\bibitem{hoge1997density}
R.~D. Hoge, R.~K. Kwan, and G.~Bruce~Pike, ``Density compensation functions for
  spiral {MRI},'' \emph{Magn. Reson. Med.}, vol.~38, no.~1, pp. 117--128, 1997.

\bibitem{kim:2018:AdaptiveRestartOptimized}
D.~Kim and J.~A. Fessler, ``Adaptive restart of the optimized gradient method
  for convex optimization,'' \emph{J Optim Theory Appl}, vol. 178, no.~1, pp.
  240--263, Jul. 2018.

\bibitem{fessler:20:omf}
J.~A. Fessler, ``Optimization methods for {MR} image reconstruction,''
  \emph{{IEEE Sig. Proc. Mag.}}, vol.~37, no.~1, pp. {33--40}, Jan. 2020.

\bibitem{noll:2005:cpm}
D.~C. Noll, J.~A. Fessler, and B.~P. Sutton, ``Conjugate phase {MRI}
  reconstruction with spatially variant sample density correction,'' \emph{IEEE
  Trans. Med. Imag.}, vol.~24, no.~3, pp. 325--336, 2005.

\bibitem{espirit}
M.~Uecker \emph{et~al.}, ``{ESPIRiT - an} eigenvalue approach to
  autocalibrating parallel {MRI:} {where} {SENSE} meets {GRAPPA},'' \emph{{Mag.
  Reson. Med.}}, vol.~71, no.~3, pp. {990--1001}, Mar. 2014.

\bibitem{fastmri}
\BIBentryALTinterwordspacing
J.~Zbontar \emph{et~al.}, ``{fastMRI: An} open dataset and benchmarks for
  accelerated {MRI},'' 2018. [Online]. Available:
  \url{http://arxiv.org/abs/1811.08839}
\BIBentrySTDinterwordspacing

\bibitem{kingma:2017:AdamMethodStochastic}
\BIBentryALTinterwordspacing
D.~P. Kingma and J.~Ba, ``Adam: {{a method}} for {{stochastic optimization}},''
  2017. [Online]. Available: \url{http://arxiv.org/abs/1412.6980}
\BIBentrySTDinterwordspacing

\bibitem{hore:2010:ImageQualityMetrics}
A.~Hor{\'e} and D.~Ziou, ``Image {{quality metrics}}: {{PSNR}} vs. {{SSIM}},''
  in \emph{{{Intl. Conf.}} on {{Patn. Recog.}} (ICPR)}, Aug. 2010, pp.
  2366--2369.

\bibitem{sparklingmrm}
C.~Lazarus \emph{et~al.}, ``{SPARKLING:} variable-density k-space filling
  curves for accelerated {T2*-weighted} {MRI},'' \emph{{Mag. Res. Med.}},
  vol.~81, no.~6, pp. {3643--61}, Jun. 2019.

\bibitem{boyer2016generation}
C.~Boyer, N.~Chauffert, P.~Ciuciu, J.~Kahn, and P.~Weiss, ``On the generation
  of sampling schemes for magnetic resonance imaging,'' \emph{SIAM J. Imaging
  Sci.}, vol.~9, no.~4, pp. 2039--2072, 2016.

\bibitem{gu:22:rlw}
H.~Gu, B.~Yaman, S.~Moeller, J.~Ellermann, K.~Ugurbil, and M.~{Akcakaya},
  ``Revisiting l1-wavelet compressed-sensing {MRI} in the era of deep
  learning,'' \emph{{Proc. Natl. Acad. Sci.}}, vol. 119, no.~33, p.
  e2201062119, Aug. 2022.

\end{thebibliography}

\appendix

This appendix analyzes the error
of approximations based on
\eqref{e,Ax} and \eqref{e,A'y},
by comparing Jacobians
computed
when \A is an exact NUDFT 
to those for an NUFFT,
denoted \Anu. 
For simplicity, the analysis is 1D,
though the conclusion extends easily to multi-dimensional NUFFTs.

The system matrix
$\A \in \CMN$ has elements
\[
\amn = \Expni{\omgm n}
,\quad
m=1,\ldots, M
,\quad
n=1,\ldots,N .
\]
Typically, an NUFFT involves three steps.
The first step applies scaling factors $s_n$
to the signal $x_n$.
The second step applies a $K$-point FFT to
the scaled signal,
where $K \geq N$
via zero-padding.
The third step interpolates
$K$ frequency locations into $M$ sampling locations of $\om$.
For efficiency,
the interpolator usually has finite support,
denoted $J > 0$.
The NUFFT
\Anu has elements as follows:
\[
\amnt =
\sum_{j=1}^{J} u_j^*(\omgm) \, s_n \, \Expni{\gamma \, (k_m + j) \, n},
\]
where $u$ denotes interpolation coefficients,
$k_m$ is an element-wise offset,
and $\gamma = 2 \pi / K$
\cite{fessler:03:nff}.

Define the NUFFT error matrix as
$\E = \Anu - \A$.
The worst-case NUFFT error has a bound
that can be written as
\[
\norminf{\E \x} \leq \veps_p \normii{\x}
,\]
where
$\veps_p$ is tabulated numerically
for various choices of
interpolation parameters $p$,
e.g.,
in \cite[Fig.~12]{fessler:03:nff}.

\comment{ 
Based on the Poisson summation formula
\cite[(4.3)]{finufft} \cite[\S V.B]{fessler:03:nff},
the error matrix has elements
\[
\emn = \frac{\sum_{l \neq 0} c_{n+lK} \, \Expni{\omgm \cdot (n+lK)}}{c_n},
\]
where $c_n$ is the inverse Fourier transform of the phase-modulated,
expanded $K$-periodic interpolator (\cite[(3.10)]{finufft}).
The maximum error is
\begin{align*}
\varepsilon_p
&\defequ
\norminf{\vect(\E)}
= \max_{m,n} |\emn|
\\&
\leq
\frac{ \max_{n,\omgm} \abs{ \sum_{l \neq 0} c_{n+lK} \, \Expni{\omgm \cdot (n+lK)} } }{ \min_{n} |c_n|}.
\end{align*}
The maximum absolute value of \emn
depends on the frequency behavior of the interpolator,
and is tabulated in figures in
\cite{fessler:03:nff}
for various NUFFT parameters $p$.
} 

The Jacobian of the forward operator \eqref{e,Ax} is
\[
\J = \frac{\partial \A\x}{\partial \om} = -\imath \, \diag{\A (\x\odot\rr)}.
\]
Let
\Jnu denote the case where an NUFFT is applied.
Since the backpropagation uses Jacobians
in the JVP calculation,
here we analyze the error of JVPs using
\J and \Jnu.
We define the worst-case relative error
for a JVP
with a (gradient) vector \v
as follows:
\begin{align*}
E_1(\om,\x,p)
&\defequ
\max_{\norminf{\v} = 1}
\norminfr{ \Jnu\v - \J\v } / \normii{\x}
\\&=
\max_{\norminf{\v} = 1}
\norminf{ ( \E \, (\x \odot \rr) ) \odot \v } / \normii{\x}
\\&=
\norminf{\E \, (\x\odot\rr) } / \normii{\x}
\\&
\leq
\varepsilon_p \normii{\x\odot\rr} / \normii{\x}
\leq \varepsilon_p \norminf{\rr}
.\end{align*}

\comment{
\begin{align*}
E_1(\om,\x,p)
&\defequ
\norminfr{ \vect(\J - \Jnu) } / \normii{\x}
\\&=
\norminf{ \vect( \diag{ \E \, (\x \odot \rr) ) } } / \normii{\x}
\\&=
\norminf{\E \, (\x\odot\rr) } / \normii{\x}
\\&
\leq
\varepsilon_p \normii{\x\odot\rr} / \normii{\x}
\\&
\leq \varepsilon_p \norminf{\rr}
.\end{align*}
}

Similarly,
the worst-case relative error of a JVP with
\eqref{e,A'y}
is bounded by
\begin{align*}
E_2(\om,\x,p)
&\defequ
\max_{\norminf{\v} = 1}
\norminf{\diag{\rr} \E' \diag{\y} \v} / \normii{\y}
\\&
\leq
\max_{\norminf{\v} = 1}
\norminf{\rr}
\norminf{\E' (\y \odot \v)} / \normii{\y}
\\&
\leq
\varepsilon_p
\norminf{\rr}
\max_{\norminf{\v} = 1}
\normii{\y \odot \v} / \normii{\y}
\\&
\leq
\varepsilon_p
\norminf{\rr}
\normii{\y} / \normii{\y}
\leq
\varepsilon_p \norminf{\rr}.
\end{align*}

\comment{ 
The relative error of \eqref{e,Ax} is defined as
$$
E(\om,\x,p) \defequ \normiir{ \J - \Jnu } / \| \x \|_2
= \normii{ \diag{ \E (\x \odot \rr) } } / \| \x \|_2
.$$
Since $\J - \Jnu$ is a diagonal matrix,
its spectral norm is 
the maximum absolute value of its
diagonal elements:
\begin{align*}
E(\om,\x,p) &=
\norminf{\E (\x\odot\rr) } / \| \x \|_2
\\&=
\max_{m} \ \abs{ [ \E (\x\odot\rr) ]_m } / \| \x \|_2
\\&
\leq
\varepsilon_p \|\x\odot\rr\|_1 / \| \x \|_2
\leq \varepsilon_p \| \rr \|_2
.\end{align*}
Similarly, the error of \eqref{e,A'y} is
bounded by
\begin{align*}
E(\om,\x,p)
&=
\max_{m} \ \abs{ [ \rr \odot (\E' \y) ]_m } / \| \y \|_2
\\&
\leq \varepsilon_p \|\y\odot\rr\|_1 / \| \y \|_2
\leq \varepsilon_p \| \rr \|_2.
\end{align*}
} 

In both cases,
the worst-case error of the NUFFT approximation
for a JVP
is bounded by the usual NUFFT error
multiplied by a constant
$\norminf{\rr}$
that is usually half of the field of view (FOV)
in imaging applications.
This constant is expected from unit analysis.
If the sampling grid $r_j$ has a unit in cm,
then the sample locations $\om$
have units in radians/cm.
Corresponding, the Jacobian matrices
in \eqref{e,Ax} and \eqref{e,A'y}
have units in cm,
because \A is unitless.
The NUFFT error $\varepsilon_p$ is unitless,
so there is an \rr-related factor
in the JVP error $E$.
In other words,
the error bounds above
depend on the choice of units.
One could express the FOV in voxels
to get the unitless error bound
$\veps_p N/2$.
%
However,
the accuracy of JVPs does not necessarily deteriorate 
with larger $N$.
Above we normalized the error by
$\normii{\x}$ or $\normii{\y}$,
whereas the Jacobians are
``scaled'' with $\normii{\x\odot\rr}$ or $\normii{\y}\normii{\rr}$.
A relative 
error could better describe
the effect on optimization.

An alternate definition uses the worst-case in the numerator
relative to an average case in the denominator,
considering the stochastic gradient descent-like optimizers.
For example, this relative error for the JVP of Jacobian \eqref{e,Ax} is
\begin{align*}
    \epsilon & \defequ \frac
    {\max_{\normii{\x} = 1} \normfror{\Jnu - \J }}
    { \sqrt{ \Epx{ \normfro{\J}^2 } }}
    = \frac
    {\max_{\normii{\x} = 1} \normii{\E \, (\x \odot \rr)}}
    {\sqrt{ \Epx{ \normii{\A \, (\x \odot \rr)}^2 } }} \\
    & \leq    \frac{\max_{\normii{\x} = 1} \sqrt{M} \norminf{\E \, (\x \odot \rr)}}
    {\sqrt{ \Epx{ \normii{\A \, (\x \odot \rr)}^2 } }}
    \leq  \frac{\sqrt{M} \veps_p \norminf{\rr}}
    {\sqrt{ \Epx{ \normii{\A \, (\x \odot \rr)}^2 } }}
,\end{align*}
where
$\Epx{\cdot}$
denotes expectation
w.r.t.
a certain distribution $p(\x)$.
For parity with the unit sphere constraint in the numerator,
we consider the case where $p(\cdot)$
is the random distribution on the unit $N$-sphere.
Use the cyclic property of the trace: 
\begin{align*}
\normii{\A \, (\x \odot \rr)}^2
&= \x' \diag{\rr} \A' \A  \diag{\rr} \x
\\&
= \trace{
 \diag{\rr} \A' \A  \diag{\rr} \x \x'
}
.
\end{align*}
Since the covariance
of random points on the $N$-sphere
is $(1/N) \I$,
the denominator's expectation is
\begin{align*}
 \Epx{ \normii{\A \, (\x \odot \rr)}^2 }
& =
\trace{ \diag{\rr} \A' \A \diag{\rr} \Epx{\x \x'}}
\\&
= \frac{1}{N} \trace{\diag{\rr} \A' \A \diag{\rr}}
\\&
= \frac{1}{N} \sum_j r_j^2 [\A'\A]_{jj}
= \frac{M}{N} \sum_j r_j^2
= \frac{M}{N} \normii{\rr}^2
.\end{align*}
Thus
we have the following bound
for the relative error:
\[
\epsilon \leq \frac
{\sqrt{M} \veps_p \norminf{\rr}}
{ \sqrt{M/N} \normii{\rr}}
=
\veps_p \sqrt{N}
\frac
{\norminf{\rr}}{\normii{\rr}}
\leq
\veps_p \sqrt{N}
.\]
Note that the bound can be tighter when considering specific formulations of \rr.
Similarly,
for the Jacobian operator \eqref{e,A'y},
the alternate error of the JVP is
\[
\epsilon 
\leq
\veps_p \sqrt{M}
.\]

\end{document}